\documentclass{article}

\usepackage{PRIMEarxiv}

\usepackage[utf8]{inputenc} % allow utf-8 input
\usepackage[T1]{fontenc}    % use 8-bit T1 fonts
\usepackage{hyperref}       % hyperlinks
\usepackage{url}            % simple URL typesetting
\usepackage{booktabs}       % professional-quality tables
\usepackage{amsfonts}       % blackboard math symbols
\usepackage{amsmath}
\usepackage{nicefrac}       % compact symbols for 1/2, etc.
\usepackage{microtype}      % microtypography
\usepackage{lipsum}
\usepackage{fancyhdr}       % header
\usepackage{graphicx}       % graphics
\usepackage{natbib}
\usepackage{eurosym}
\usepackage{booktabs,tabularx}
\usepackage{rotating}
\usepackage{pdflscape}
\usepackage{graphicx}
\usepackage{subcaption}
\usepackage{placeins}        % for float barrier

\usepackage{algorithm}
\usepackage{algpseudocode}

\DeclareMathOperator*{\argmax}{arg\,max}

\usepackage{todonotes}
\let\opentask\todo % redefining todo{} as todo[inline]{} to appear in the text not in margins
\renewcommand{\todo}[1]{\opentask[inline,color=red!40]{#1}}
\title{Predictive Vehicle Repositioning for On-Demand Ride-Pooling Services
}

\author{
  Roman Engelhardt\\
  Chair of Traffic Engineering and Control \\
  Technical University of Munich \\
  Arcisstraße 21 \\
  80333 Munich, Germany \\
  \texttt{Corresponding Author: roman.engelhardt@tum.de} \\
  \And
  Hani S. Mahmassani \\
  Department of Civil and Environmental Engineering \\
  Northwestern University \\
  Evanston, IL 60208 \\
  United States \\
  \texttt{masmah@northwestern.edu} \\
  \And
  Klaus Bogenberger \\
  Chair of Traffic Engineering and Control \\
  Technical University of Munich \\
  Arcisstraße 21 \\
  80333 Munich, Germany \\
  \texttt{klaus.bogenberger@tum.de} \\
}

\begin{document}
\maketitle

\begin{abstract}
On-Demand Ride-Pooling services have the potential to increase traffic efficiency compared to private vehicle trips by decreasing parking space needed and increasing vehicle occupancy due to higher vehicle utilization and shared trips, respectively. Thereby, an operator controls a fleet of vehicles that serve requested trips on-demand while trips can be shared. In this highly dynamic and stochastic setting, assymetric spatio-temporal request distributions can drive the system towards an imbalance between demand and supply when vehicles end up in regions with low demand. This imbalance would lead to low fleet utilization and high customer waiting times. This study proposes a novel rebalancing algorithm to predictively reposition idle fleet vehicles to reduce this imbalance. The algorithm first samples artificial requests from a predicted demand distribution and simulates future fleet states to identify supply shortages. An assignment problem is formulated that assigns repositioning trips considering multiple samples and forecast horizons. The algorithm is implemented in an agent-based simulation framework and compared to multiple state-of-the-art rebalancing algorithms. A case study for Chicago, Illinois shows the benefits of applying the repositioning strategy by increasing service rate and vehicle revenue hours by roughly 50\% compared to a service without repositioning. It additionally outperforms all comparison algorithms by serving more customers, increasing the pooling efficiency and decreasing customer waiting time regardless of the forecasting method applied. As a trade-off, the computational time increases, but with a termination within a couple of seconds it is still applicable for large-scale real world instances.
\end{abstract}

% keywords can be removed

\section{Introduction}
By 2050, the UN projects that 68\% of the world’s population will live in urban areas, up from 55\% in 2018~\citep{UnitedNations.2018}. This trend is accompanied by an increase in travel demand, traffic congestion, air pollution and noise. While Covid-19 has reduced congestion in 2020, pre-pandemic levels have again been reached in many US and European cities~\citep{TomTom.2022}. The transportation sector is additionally responsible for a large share of greenhouse gas emissions, 28\% in the US~\citep{USEPA.2021} and 20\% in Germany~\citep{Umweltbundesamt.2022}. It is therefore imperative to enhance traffic efficiency in order to combat resource scarcity and climate change while providing for the mobility needs of people and goods. 

In contrast to the inefficient usage of private vehicles characterized by low occupancy and utilization rates, on-demand ride-pooling (ODRP) services have emerged as a promising solution to enhance efficiency while maintaining a comparable level of service as private vehicles. In ODRP services, customers request trips on-demand, and an operator dynamically assigns schedules to its vehicles to serve the requested trips. This system allows multiple customers to share their trip. The potential replacement of private vehicle trips by ODRP services holds the key to significantly reducing the number of vehicles in urban areas and increasing overall vehicle occupancy. With the imminent prospect of autonomous vehicles and their low operating costs, ODRP services can be offered at affordable fares~\citep{Boesch.2016}.

Recent studies have focused on quantifying the potential benefits of ODRP services. For instance, as simulation study~\citep{AlonsoMora.2017} showed that just 3000 vehicles could efficiently serve the taxi demand in New York City.  Another study~\citep{D.Fiedler.2018} revealed that network congestion could be drastically reduced if all private vehicle trips were replaced by an ODRP service in Prague. However, these studies assumed a relatively high market penetration, exceeding 100k trips per day, which facilitates finding shareable trips. Research by~\citep{Engelhardt.2019} and~\citep{Fagnant.2018} highlighted the importance of trip density for ride-pooling to overcome empty pick-up trips through effective ride-sharing. This scaling property of ride-pooling has been further corroborated by macroscopic~\citep{Tachet.2017, Bilali.2020} and graph-based~\citep{Santi.2014, Kucharski.2020} studies.

The control problem underlying an ODRP service can be represented as a Dial-a-Ride Problem (DARP). The DARP has been a subject of study for over four decades (e.g., by~\citep{Cordeau.2003, Cordeau.2006}). However, it is known to be NP-hard, limiting exact solution methods to small system sizes. Nevertheless, the solution algorithms must be capable of providing short runtimes to accommodate new customers on-demand. Consequently, various heuristic methods have been developed to meet these criteria. Some of these algorithms are based on insertion heuristics~\citep{Jaw.1986, S.Ma.2013}, meta-heuristics~\citep{Jung.2016, Massobrio.2016}, column generation~\citep{Riley.2019} and graph-based approaches~\citep{AlonsoMora.2017, Engelhardt.2020, Simonetto.2019}.
These heuristic assignment formulations often share a common limitation of disregarding future information during the assignment process. This can result in spatio-temporal imbalances of vehicles when the demand distribution is asymmetric. In such cases, vehicles may accumulate in regions with low demand while being scarce in areas with high demand, leading to elevated rejection rates or prolonged waiting times for customers.

To address this issue, this study proposes an algorithm designed to distribute idle vehicles strategically within the operating area of an ODRP service to predictively accommodate future demand. This problem is referred to as the "repositioning (or rebalancing) problem”. By sampling from a forecast distribution, this algorithm is designed to consider ridesharing in its formulation. 

In the next section, a literature review for solution methods of the rebalancing problem is provided, followed by a contribution statement of this study. The study then elaborates on the methodology applied in detail, which is subsequently tested through a case study based on Chicago's TNC (Transportation Network Companies) data. After presenting the results for the case study, the paper is concluded with a summary and key takeaways.

\section{Literature Review}
The rebalancing problem arises predominantly in systems characterized by high dynamism and stochasticity. Just reacting myopically to incoming demand will lead to an imbalanced system when the spatio-temporal demand patterns are not symmetric. These features are particularly common in mobility-on-demand (MoD) services but have been also studied in the area of disaster response (e.g.~\citep{Gao.2022}) or ambulance rebalancing (e.g.~\citep{Brotcorne.2003}). Concerning   the underlying rebalancing problem, these different domains of application can mainly be distinguished by the frequency with which repositioning strategies are employed. This frequency  is  mainly defined by the cost for rebalancing, the time scale that drives the imbalance and constraints like available staff to perform the trips.

In MoD services, dedicated drivers are available for each vehicle (or vehicles can rebalance themselves when automated vehicles are considered). In such scenarios, the cost of rebalancing is relatively low, and decisions for rebalancing can be made continuously.  Unlike  ride-sourcing services like Uber or Lyft, where drivers tend to reposition themselves in a greedy manner to maximize their revenue~\citep{Castillo.2017}, this study considers an MoD service that is centrally controlled by an operator. The operator assigns repositioning trips to optimize the overall fleet performance instead of the profit of single drivers.

A common approach for the ride-hailing use case, which does not allow for shared trips, is to aggregate the demand forecast into zones. Since each anticipated future trip requires precisely one vehicle as supply, analytical approximations for zonal demand-supply imbalances can be formulated. For instance, \cite{Zhang.2016} used a queuing theoretical approach to formulate the resulting rebalancing problem to stabilize a Jackson Network. On the other hand, \cite{Valadkhani.2023} proposes a macroscopic model to predict future fleet states and rebalance vehicles accordingly to optimize profit. \cite{Dandl.2019b} evaluated the impact of spatio-temporal demand forecast aggregation and found that less aggregated demand profits the ride-hailing service. However, it is crucial to find an appropriate balance, as overly small zones may cause the approximated spatial coverage of vehicles to extend beyond the zone boundaries. To reduce the impact of the spatial aggregation method for rebalancing, \cite{Syed.2021} therefore introduced spatial correlations based on Gaussian Kernels between zones, while \cite{Zhu.2022} approximate the spatial supply density by Voronoi cells originating from each vehicle. 

When trips can be shared in ODRP services, the problem complexity increases, as the relationship between expected demand and required supply becomes non-trivial. Some studies have suggested methods to address this challenge: \cite{Wallar.2018} introduced a linear scaling factor of predicted demand to convert expected demand to supply, allowing to use a computationally efficient macroscopic model. Alternatively, \cite{Schlenther.2023} proposed aligning relative demand and supply distributions instead of rebalancing vehicles to absolute measures of demand. \cite{Tsao.2019} proposed a model predictive control approach to steer vehicles towards future expected demand, but this method is limited to a maximum of two requests sharing a trip, and the case study is highly aggregated with only up to 25 zones. \cite{Sayarshad.2017} formulated a rebalancing problem based on Markov Decision Processes, but the problem size is restricted to 6 zones in their case study. As analytical formulations are hard to find, multiple studies proposed deep learning approaches which show promising results~\citep{ChengLi.2022, Gueriau.2020, Wen.2017, Chouaki.2022}.

Another approach to estimate future supply shortages involves sampling requests from a forecast distribution and using them to compute possible future vehicle routes. This approach allows for the direct inclusion of design parameters, such as time constraints and objective functions, to construct the routes and synchronize assignment and rebalancing. However, sampling methods can be computationally demanding, as they require solving vehicle routing problems. \cite{Li.2019} proposed a solution method for the stochastic DARP using sampling, but the problem size was restricted to 4 vehicles. A large-cale rebalancing method has been developed by \cite{AlonsoMora.2017b}: Samples from future requests are directly included in the assignment algorithm. While this method showed promise in large-scale simulations for Manhattan, the inclusion of future request samples drastically increased computational time, necessitating the addition of multiple time-outs in the assignment process to manage computational demands effectively.

\section{Contributions}
This study presents a novel rebalancing algorithm tailored specifically for ride-pooling services, taking into account the potential for trip sharing when calculating future supply shortages. The algorithm is designed based on sampling requests from expected trip distributions, but still achieves termination within a couple of seconds enabling its application to large-scale instances with hundreds of vehicles.  This efficiency makes the algorithm suitable for large-scale real-world implementation.

The proposed algorithm is implemented within an agent-based simulation framework. A case study is conducted using data from Chicago, Illinois, to quantify the benefits of rebalancing and benchmark the performance of this algorithm against other state-of-the-art rebalancing algorithms.

\section{Methods}
This study assumes an operator of an ODRP service that controls a fleet of vehicles $v \in V$. Over time, customers request trips from the operator. The operator centrally controls its vehicles, i.e. the operator performs actions $A_t$ in certain time steps $t$ depending on the current system state $S_t$. The goal is to perform actions (i.e. assign tasks to its vehicles) that optimize the operator’s long-term profit $P$. The control problem can be formulated as
\begin{align}
    \max_A P \qquad & \qquad P = \sum_t P_t(A_t, S_t) & \\
    \text{s.t.:} \qquad & \qquad S_{t+1} = \Omega(S_t, A_t, s_{t+1}) & \forall t
\end{align}

$P_t$ evaluates the profit generated at time-step $t$, while $\Omega$ is a state transition function defining the system evolution based on performed action and exogenous variables $s_{t+1}$ that describe state changes independent of operator actions, e.g. new customers requesting trips.

If stochastic information about future exogenous state changes is available, the Bellman equation can be used to evaluate optimal actions $A_t^*$ in theory:
\begin{align}
    \label{eq:bellmann}
    & A_t^* = \argmax_{A_t} (P_t(A_t, S_t) + \mathbb{E}[\sum_t \gamma^T P_{t+1}(A_{t+1}, S_{t+1})]) & \\
    \text{s.t.:} \qquad & \qquad S_{t+1} = \Omega(S_t, A_t, s_{t+1}) & \forall t
\end{align}

The second term evaluates expected future rewards while and weights them by the parameter $\gamma \in [0, 1]$.

In theory, Monte Carlo simulations can be used to estimate future rewards and dynamic programming approaches can be applied to determine the optimal actions $A_t^*$. Nevertheless, this solution method would require solving a lot of DARPs, which is computationally not tractable for large-scale ODRP systems.A common approach is to separate fleet operator actions into two sequential steps: 
\begin{enumerate}
    \item \textbf{Assignment}: In this step, the fleet operator reacts to new customer trip requests and updates the schedules of vehicles to efficiently serve these requests.
    \item \textbf{Repositioning}: The focus of this study is on the repositioning step. In this phase, future expected request distributions are evaluated to match spatio-temporal demand and supply distribution to optimize future service rate and vehicle utilization.
\end{enumerate}

In the following, the simulation framework the control algorithm is implemented in is introduced, followed by a high level description of the applied assignment algorithm. Finally, the proposed rebalancing algorithm is formulated in detail.

\subsection{High-Level Framework Description}
This study utilizes the open-source agent-based simulation framework FleetPy~\citep{TUMVT.2022} which focuses on the simulation of MoD services~\citep{Engelhardt.2022}. In the simulation, customers request trips from an ODRP operator, which, in turn, assigns schedules to its fleet vehicles to fulfill the requests. Vehicles travel in a network $G=(N,E)$ with nodes $n \in N$ and edges $e \in E$. Each customer request $r_i$ is described by a tuple of origin location $o_i \in N$, destination location $d_i \in N$ and the time of the request $t_i$. Customers expect to be picked-up as soon as possible and are considered impatient, meaning that if the service cannot be provided within a maximum waiting time of $t_{max}^{wait}$, they will not use the service. Additionally, customers are willing to accept a detour for pooling of up to $\Delta_{max}^{det}$ relative to the direct travel time from their origin to their destination. The fleet of the operator consist of vehicles $v \in V$ with fleet size $|V|$. Each vehicle has a capacity of $c_v$ passengers.
The operator assigns schedules $\psi$, i.e. list of stops where customers can board or alight the vehicle. Between stops, vehicles drive on the fastest route in the network. Schedules are considered feasible if
\begin{enumerate}
    \item each customer served by the schedule is dropped off after being picked up,
    \item for each customer, the maximum waiting time constraint $t_{max}^{wait}$ and maximum travel time constraint $\Delta_{max}^{det}$ is fulfilled,
    \item at no time, more than $c_v$ passengers are in the vehicle.
\end{enumerate}
The operator rates feasible schedules by the objective function
\begin{align}
\label{eq:obj}
    \rho{\psi} = \tau(\psi) - \pi |R_\psi| ~ .
\end{align}
$\tau(\psi)$ measures the time needed to fulfill the schedule (system time), while $|R_\psi|$ refers to the number of customers that are served by it. $\pi$ is a positive, sufficiently large assignment reward to prioritize serving requests over minimizing system time, when this objective function is to be minimized. It's important to note that customer-centric terms, such as minimizing waiting and travel time for individual customers, could be included in the objective function as well. However, this study neglects these terms, as it is argued that the operator has already applied sufficiently tight time constraints for pick-up and drop-off times. This allows the operator to fully focus on assigning efficient shared routes within the time constraints applied.

\subsection{Assignment}

The assignment algorithm assigns requests in batches every $\Delta t_A = 60$s (other values may of course be used). The algorithm applied is based on the one proposed by~\cite{AlonsoMora.2017}. As it is not the focus of this study, it is only described on a high level, while interested readers are referred to~\cite{Engelhardt.2020} for details of the implementation.

The idea of the algorithm is to first compute all feasible schedules based on current fleet states and active customers (customers waiting for pick-up and customers on-board of a vehicle) and solve an Integer Linear Problem (ILP) to assign schedules to vehicles in a second step. To compute all feasible schedules, a guided search is applied that explicitly exploits time constraints for customer pick-up and drop-off.

A vehicle-to-request-bundle (V2RB) $\Psi(v, R_\Psi)$ is defined as the collection of all feasible schedules that serve the same set of requests $R_\Psi$. The feasible schedule with the minimum objective based on Equation~\ref{eq:obj} represents the V2RB and the objective of the V2RB. The grade of a V2RB is defined as the number of requests that are served by the V2RB. Three main conditions are required to hold for the existence of a V2RB:
\begin{enumerate}
    \item A V2RB of grade 1 can only exist, if the vehicle can reach the one request before the maximum waiting time elapsed.
    \item A V2RB of grade 2 can only exist, if there is a feasible schedule of a hypothetical vehicle serving both requests starting at the origin of one of the two requests.
    \item A V2RB serving the requests $R_\Psi$ of grade n can only exist if all V2RBs of grade $n-1$ exist, that serve a subset of $R_\Psi$. E.g. if a feasible schedule serving $(r_1, r_2, r_3)$ exists, also feasible schedules that only serve $(r_1, r_2)$, $(r_2, r_3)$ and $(r_1, r_3)$ have to exist.
\end{enumerate}
These three conditions allow computing all feasible schedules by gradually increasing the grades of V2RBs. New V2RBs are created by inserting new requests into the schedules of lower-grade V2RBs.

Once all V2RBs are created, the following ILP is solved to assign V2RBs (its representative schedule) to vehicles:
\begin{align}
    \label{eq:ass_1}
    \text{Minimize:} \qquad & \qquad \sum_{v \in V} \sum_{m \in \Omega_v} \rho_{v, m} z_{v, m} & \\
    \label{eq:ass_2}
    \text{s.t.:} \qquad & \qquad \sum_{v \in \Omega_v} z_{v, m} \leq 1 & \forall v \in V \\
    \label{eq:ass_3}
    & \qquad \sum_{v \in V} \sum_{m \in \Omega_v^i} z_{v, m} \leq 1 & \forall r_i \in R_u \\
    \label{eq:ass_4}
    & \qquad \sum_{v \in V} \sum_{m \in \Omega_v^i} z_{v, m} = 1 & \forall r_i \in R_a \\
    \label{eq:ass_5}
    & \qquad z_{v, m} \in \{0, 1\}
\end{align}
$\Omega_v$ refers to the set of V2RBs for vehicle $v$. $\rho_{v, m}$ is the objective value of the $m$-the V2RB served by vehicle $v$. $z_{v, m}$ is a binary decision variable to assign V2RBs to vehicles. Equation~\ref{eq:ass_2} ensures that maximally one V2RB is assigned to each vehicle. Equation~\ref{eq:ass_3} ensures that yet unassigned requests in the set $R_u$ are assigned maximally once. $\Omega_v^i$ thereby refers to the set of V2RBs of vehicle $v$ that include request $r_i$. Similarly, Equation~\ref{eq:ass_4} ensures that previously assigned requests ($R_a$) are assigned again.

\subsection{Rebalancing}

Once a vehicle completes a schedule, it would only be assigned to a new schedule if a trip request is made in the vicinity of a maximum driving time of $t_{max}^{wait}$. To avoid vehicles being stuck in regions where fewer requests are made than vehicles arrive, vehicles need to be repositioned to regions with undersupply to increase service availability and vehicle utilization.  Calculating rebalancing trips usually requires three main steps:
\begin{enumerate}
    \item A forecast of future demand. This demand is often aggregated on a zonal level within certain time intervals.
    \item A methodology to estimate expected profit for sending vehicles to a specific zone or an expected number of required vehicles.
    \item An algorithm to assign repositioning trips for idle vehicles to specific zones.
\end{enumerate}
This study focuses on the last two steps. It assumes the ODRP operating area is partitioned into zones $Z$. A demand forecast is available estimating the expected number of customers $\lambda_{i,j}^T$ requesting trips from zone $i \in Z$ to zone $j \in Z$ within a time window between $[T, T+\delta_T]$ . 

\subsubsection{General Idea}

The proposed algorithm follows a sampling approach to address future vehicle imbalances and make informed decisions. The rebalancing algorithm is applied less frequently than the assignment algorithm in steps of $\delta_T = 900$s. By sampling artificial requests from a forecast distribution, the algorithm generates actual routes that accurately consider service design parameters also applied in the assignment algorithms. Thereby, it can estimate the number of customers that can be served by the same idle vehicle while also considering the capacity of currently en-route vehicles to accommodate future requests. As an output, the idle vehicles are sent towards the locations of the expected first pick-ups. The en-route vehicles remain following their original V2RBs.

\begin{figure}[!ht]
  \centering
  \resizebox{0.95\textwidth}{!}{\includegraphics{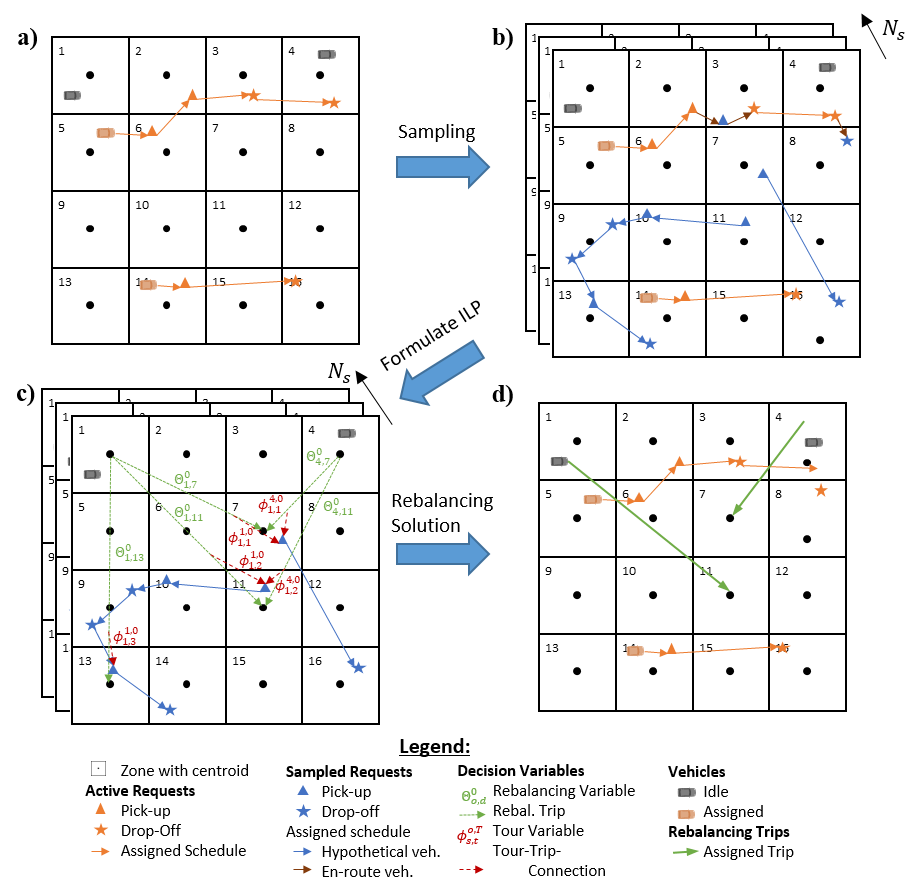}}
  \caption{Example sketch for solving the rebalancing problem for ride-pooling.}\label{fig:sketch_alg}
\end{figure}

Figure~\ref{fig:sketch_alg} presents an overview of the rebalancing algorithm. In the first step (a), the algorithm takes as input only all currently en-route (not idle)  vehicles and their assigned schedule, which are used to estimate their ability to accommodate future requests. For $N_S$ different samples (b), future requests are drawn from the forecast distribution defined by $\lambda_{i,j}^T$ within a forecast horizon $\mathfrak{H}$, covering all temporal forecast bins $T \in \{t, t+\delta_t, ..., t + \mathfrak{H} \}$. For each sample, future vehicle states are simulated to identify supply shortages. Requests that cannot be accommodated by en-route vehicles form a new schedule  for a hypothetical vehicle available starting in the corresponding zone of the request’s origin. Each hypothetical vehicle represents an actual idle vehicle that would need to be repositioned to the corresponding zone. A zone-based assignment problem is formulated (c) that assigns idle vehicles to reposition to the zone of hypothetical vehicles (d). In the following paragraphs, the sampling process and the assignment problem is described in detail. 

\subsubsection{Sampling Future Fleet States}

The algorithm to compute future vehicle states is sketched in Algorithm~\ref{alg:sampling}. Input to the algorithm are currently en-route vehicles with their assigned schedules and the forecast distribution described by $\lambda_{i,j}^T$ with forecast horizon $\mathfrak{H}$. $N_S$  different request samples are created to reduce stochastic variance. A Poisson process with rate $\lambda_{i,j}^T$ determines the number of trips requested from zone $i$ to zone $j$. A random node from zone $i$ and zone $j$ is drawn as origin and destination of the request, respectively. The request time is randomly chosen within the time interval $[T, T+\delta_T]$. 

Fleet states are progressed into the future in time steps of 60s. Each time step, the assignment of new requests is treated at first. As the rebalancing time step $\delta_T$ is generally smaller than the forecast horizon H, it is crucial that the request assignment is computationally efficient. Performing the previously described assignment algorithm can be computationally too costly to be applied in the rebalancing step. Therefore, an insertion heuristic~\citep{Jaw.1986} is used to find feasible schedules for the request: The request is only inserted into the currently assigned schedule of each vehicle that can reach the origin of the request within $t_{max}^{wait}$. The resulting vehicle schedule that minimizes the objective of Equation~\ref{eq:obj} is assigned to the vehicle. If no solution is found, a new hypothetical vehicle is created at the zone centroid of the request origin and assigned to serve the request. After all sampled requests of the time step are assigned, vehicles are moved according to their assigned schedule. 

After all sampled requests are addressed, input parameters for the rebalancing formulation are constructed. The start zone $o_{s,t}$ of each hypothetical vehicle marks a possible future supply shortage. The objective value $\rho_{s,t}$ of the created schedule computed with Equation~\ref{eq:obj} and estimates the operator profit for providing an idle vehicle at this location. The starting time $\tau_{s,t}$ of the schedule estimates the latest arrival time of a vehicle in this zone to serve this schedule. It might not always be possible to find idle vehicles to reach the zone in time. In this case, it might be useful for an idle vehicle to enter the hypothetical vehicle’s schedule at a later time and location. Therefore, sub-schedules are defined for each hypothetical vehicle’s schedule: At each stop, the algorithm checks whether the vehicle occupancy of the schedule would be zero. If this is the case, a new sub-schedule is created. Similarly, for each sub-schedule the $o_{s,t}$, $\tau_{s,t}$ and $\rho_{s,t}$ is computed.  

\begin{algorithm}
\caption{Creating Future Schedules From Sampled Requests}\label{alg:sampling}
\begin{algorithmic}

\Require Assigned vehicles with current schedules, forecast distribution $\lambda_{i,j}^T$
\Ensure List of start\_zone, start\_time, objective, sub\_tour\_index, tour\_index, sample

\State $V_A \gets$ Assigned vehicles with current schedules
\State $V_R \gets$ Empty list of new rebalancing vehicles with schedules
\State $T \gets$ Empty list start\_zone, start\_time, objective, sub\_tour\_index, tour\_index, sample
\State $s \gets 0$
\For{$N_S$ samples}
    \State $request\_sample \gets$ Sample requests from $\lambda_{i,j}^T$
    \ForAll{time steps}
        \ForAll{$sampled\_requests$ in time step}
            \State $best\_schedule \gets $ None
            \ForAll{$vehicles$ with $schedule$ in $V_A + V_R$}
                \State $new\_schedule$ $\gets$ insert($sampled\_request$, $schedule$)
                \If{$objective(best\_schedule) < objective(new\_schedule)$}
                    \State $best\_schedule \gets new\_schedule$
                \EndIf
            \EndFor
            \If{$best\_schedule$ is not None}
                \State update schedule of corresponding vehicles
            \Else 
                \State create new artificial vehicle at origin of request and add to $V_R$
            \EndIf
        \EndFor
        \State move vehicles in $V_A + V_R$ according to assigned schedules
    \EndFor
    \State $u \gets 0$
    \ForAll{$vehicles$ with $schedule$ in $V_R$}
        \State $t \gets 0$
        \ForAll{$stop$ in $schedule$ with zero vehicle occupancy}
            \State $sub\_schedule \gets $ remove preceding stops from $schedule$
            \State $o_{s,t} \gets start\_zone(sub\_schedule)$
            \State $\tau_{s,t} \gets start\_time(sub\_schedule)$
            \State $\rho_{s,t} \gets objective(sub\_schedule)$
            \State add $(o_{s,t}, \tau_{s,t}, \rho_{s,t}, t, u, s)$ to $T$
            \State $t \gets t+1$
        \EndFor
        \State $u \gets u+1$
    \EndFor
    \State $s \gets s+1$
\EndFor

\end{algorithmic}
\end{algorithm}

\subsubsection{Rebalancing Formulation}

An ILP is formulated to assign rebalancing trips to idle vehicles to serve the sampled schedules. Idle vehicles are aggregated on a zonal level to decide for rebalancing trips between zone o and d. As the forecast horizon H is considered larger than the reposition period $\delta_T$ also possible future rebalancing trips are considered. The decision variable $\theta_{o,d}^0$ refers to immediate rebalancing actions that are performed after the problem is solved. $\tilde{\theta}_{o,d}^{T,s}$) on the other hand refers to potential future rebalancing trips in time step $T \in \{0, 1, 2, ..., T_{max} = \frac{\mathfrak{H}}{\delta_T} \}$ in sample $s$. Depending on how the real system evolves, they might or might not be realized at a later time. Note that the immediate rebalancing decision variable $\theta_{o,d}^0$ is independent of the sample s as only one decision can be made, which should lead to a good performance across all possible realizations sampled. The optimization problem is defined as follows:

\begin{align}
    \nonumber
    \text{Minimize:}  & \qquad \qquad \sum_{o,d \in Z} \left( c_{o,d} \theta_{o,d}^0 + \frac{1}{N_S} \sum_{s=0}^{N_S} \sum_{T=1}^{T_{max}} \gamma^T c_{o,d} \tilde{\theta}_{o,d}^{T,s} \right) + & \\
    \label{eq:form_obj}
    & \qquad \qquad \qquad \qquad \qquad + \frac{1}{N_S} \sum_{s=0}^{N_S} \sum_{t \in T(s)} \sum_{(d,T) \in A(t)} \gamma^T \rho_{s,t} \phi_{s,t}^{d,T} & \\
    \label{eq:form_1}
    \text{s.t.:} & \qquad  \sum_{d \in Z} \theta_{o,d}^0 \leq V_0^{idle}   & \forall o \in Z  \\
    \nonumber
    & \qquad \sum_{d \in Z} \tilde{\theta}_{o,d}^{T,s} \leq V_0^{idle} - \sum_{o \in Z} \theta_{o,d}^0 + & \\
    \label{eq:form_2}
    & \qquad \qquad \qquad + \sum_{\tau = 1}^{T-1} \left(\Delta V_{\tau, s, o}^{idle} + \sum_{d, t, \in D(o, \tau)} \phi_{s,t}^{d, \tau} - \sum_{o \in Z} \tilde{\theta}_{o,d}^{\tau,s} \right) & \forall o \in Z, \forall s \in N_s, \forall T \neq 0 \\
    \label{eq:form_3}
    & \qquad \theta_{o,d}^0 = \sum_{t \in (s, d, T)} \phi_{s,t}^{d, T} & \forall o,d \in Z, \forall s \\
    \label{eq:form_4}
    & \qquad \tilde{\theta}_{o,d}^{\tau,s} = \sum_{t \in (s, d, T)} \phi_{s,t}^{d, T} & \forall o,d \in Z, \forall s, \forall T \in [1,..., T_{max}] \\
    \label{eq:form_5}
    & \qquad \sum_{t \in U_{\kappa}(s)} \sum_{o, T} \leq 1 & \forall s, \forall U_\kappa (s) \\
    \label{eq:form_6}
    & \qquad \theta_{o,d}^0, \tilde{\theta}_{o,d}^{T,s} \in \mathbb{N}_0^+ & \forall o, d, T, s \\
    \label{eq:form_7}
    & \qquad \phi_{s,t}^{d, T} \in \{0, 1\} & \forall d, T, s, t
\end{align}

The first line of the objective in Equation~\ref{eq:form_obj} reflects the trade-off between costs and expected profit for repositioning. $c_{o,d} \geq 0$  are the costs (the travel time between the corresponding zone centroids). The factor $\gamma \in [0,1]$ weights the costs for assigning future rebalancing trips in line with the Bellmann Equations (Equation~\ref{eq:bellmann}). The first term in the first line considers immediate rebalancing decisions, while the second term considers future ones. The second line in the objective function reflects expected profit from rebalancing trips. $\rho_{s,t} \leq 0$ is the objective value calculated in the sampling process for assigning trip $t$ from sample $s$. $\phi_{s,t}^{d,T}$ is the corresponding decision variable: It takes the value 1 if a rebalancing trip from zone $d \in Z$ in time step $T$ is assigned to trip $t$ from sample $s$. The set $T(s)$ reflects all tours sampled in $s$, while the set $A(t)$ collects all possible rebalancing trips that can reach the tour $t$ in time. Equation~\ref{eq:form_1} and Equation~\ref{eq:form_2} constrain the number of vehicles that can be rebalanced per zone $o \in Z$. While for immediate rebalancing trips in Equation~\ref{eq:form_1}, only the number of currently idle vehicles per zone $V_o^{idle}$ need to be considered, future rebalancing trips in Equation~\ref{eq:form_2} also considers that vehicles already have been rebalanced out of the zone in previous  decision time steps, new vehicles with current assignments become idle ($\Delta V_{\tau, s, o}^{idle}$), or vehicles become idle after they finished their assigned tour after the rebalancing trip. $D(o,\tau)$ thereby is the set of tours that are finished in zone $o$ and decision period $\tau$. The Equations~\ref{eq:form_3} and \ref{eq:form_4} relate rebalancing trips and the assignment of corresponding sampled tours. Note that in Equation~\ref{eq:form_3} the decision variable is not indexed by the sample $s$, i.e. immediate rebalancing trips can be assigned to multiple tours, one per sample. With this constraint, efficient decisions for immediate rebalancing trips across all samples are made. In contrast, future rebalancing trips in Equation~\ref{eq:form_4} are different for each sample. Equation~\ref{eq:form_5} ensures that each tour is assigned only once. Finally, Equation~\ref{eq:form_6} and~\ref{eq:form_7} define rebalancing trips and tour assignment variables as integer and binary variables, respectively.

\subsection{Rebalancing - Comparison Algorithms}

To evaluate the performance of the proposed rebalancing algorithms, it is compared with other algorithms from the literature as a benchmark that shall be introduced on a high level.

\subsubsection{No Rebalancing}

No rebalancing is applied.

\subsubsection{Reactive Rebalancing \textit{React}}

This algorithm is described in \cite{AlonsoMora.2017} and is based on an expected autocorrelation of demand. After each assignment step, the locations of unserved requests are tracked. Anticipating future demand at these locations, available idle vehicles are rebalanced there by solving an assignment problem, minimizing the overall travel time. Alongside its simplicity, the advantage of this algorithm is that no forecast for future demand is necessary.  

\subsubsection{Queuing Theoretical Rebalancint \textit{QT}}

This problem formulation uses queuing theoretical considerations to stabilize a Jackson network \citep{Zhang.2016}. The assignment problem to be solved can be formulated as

\begin{align}
    \label{eq:qt_1}
    \text{Minimize:} \qquad & \qquad \sum_{o,d \in Z } \tau_{o,d} \beta_{o,d} & \\
    \label{eq:qu_2}
    \text{s.t.:} \qquad & \qquad \sum_{d \neq o} (\beta_{o,d} - \beta_{d,o}) = - \mu_{QT} \sum_{d \neq o} (\lambda_{o,d} - \lambda_{d,o}) - I_o + \sum_d \frac{I_d}{|Z|} & \forall d \in Z \\
    \label{eq:qu_3}
    & \qquad \beta_{o,d} \geq 0 & \forall o,d \in Z
\end{align}

$\beta_{o,d}$ is the (non-integer) decision variable to rebalance vehicle from o to d while $\tau_{o,d}$ is the interzonal travel time. The constraint of Equation~\ref{eq:qu_2} balances the expected in- and out-flow of each zone. $\lambda_{o,d}$ are the expected number of trip requests between zones $o$ and $d$ within a forecast horizon $\mathfrak{H}_{QT}$. $I_d$ are the number of idle vehicles per zone. The last two terms try to distribute remaining idle vehicles evenly across zones. $\mu_{QT}$ is a demand scaling factor, introduced in this study to consider sharing of trips.

To assign vehicles, the value $\beta_{o,d}$ is rounded to the next integer after the problem is solved. Additionally, this formulation does not constrain the number of assigned vehicles to be smaller or equal the number of idle vehicles. Therefore, for each origin zone, the assignment of idle vehicles is performed in random order, and stops, if no idle vehicle remains in a zone.

\subsubsection{Horizon-base Rebalancing \textit{Hor}}
This algorithm is proposed in \cite{Wallar.2018} and considers the time when rebalancing vehicles arrived in their target zone. It is formulated as

\begin{align}
    \label{eq:hor_1}
    \text{Minimize:} \qquad & \qquad \sum_{o,d \in Z } (\mathfrak{H}_{Hor} - \tau_{o,d}) \lambda_{o,d} \beta_{o,d} & \\
    \label{eq:hor_2}
    \text{s.t.:} \qquad & \qquad \sum_{d \in Z} \beta_{o,d} \leq I_o    & \forall o \in Z \\
    \label{eq:hor_3}
    & \qquad \beta_{o,d} (\mathfrak{H}_{Hor} - \tau_{o,d}) \geq 0 & \forall o,d \in Z \\
    \label{eq:hor_4}
    & \qquad \sum_{o \in Z} \beta_{o,d} (1 - \frac{\tau_{o,d}}{\mathfrak{H}_{Hor}} ) \leq \lambda_d \mu_{Hor} & \forall d \in Z
\end{align}

$\mathfrak{H}_{Hor}$ is the forecast horizon applied for this strategy, while $\lambda_d$ expected number of requests arriving in zone $d$ during the forecast horizon. Equation~\ref{eq:hor_2} constrains the number of vehicles that can be rebalanced, Equation~\ref{eq:hor_3} ensures that vehicles reach the rebalancing destination within the horizon and Equation~\ref{eq:hor_4} constrains the supply in target  zones. The left-hand side computes the number of vehicles rebalancing to the zone weighted by time they are available in this zone. The right-hand side estimates the expected demand for vehicles. $\mu_{Hor}$ is a scaling factor to specify an acceptable level of oversaturation.

\section{Case Study}

\begin{figure}[!ht]
  \centering
  \resizebox{0.7\textwidth}{!}{\includegraphics{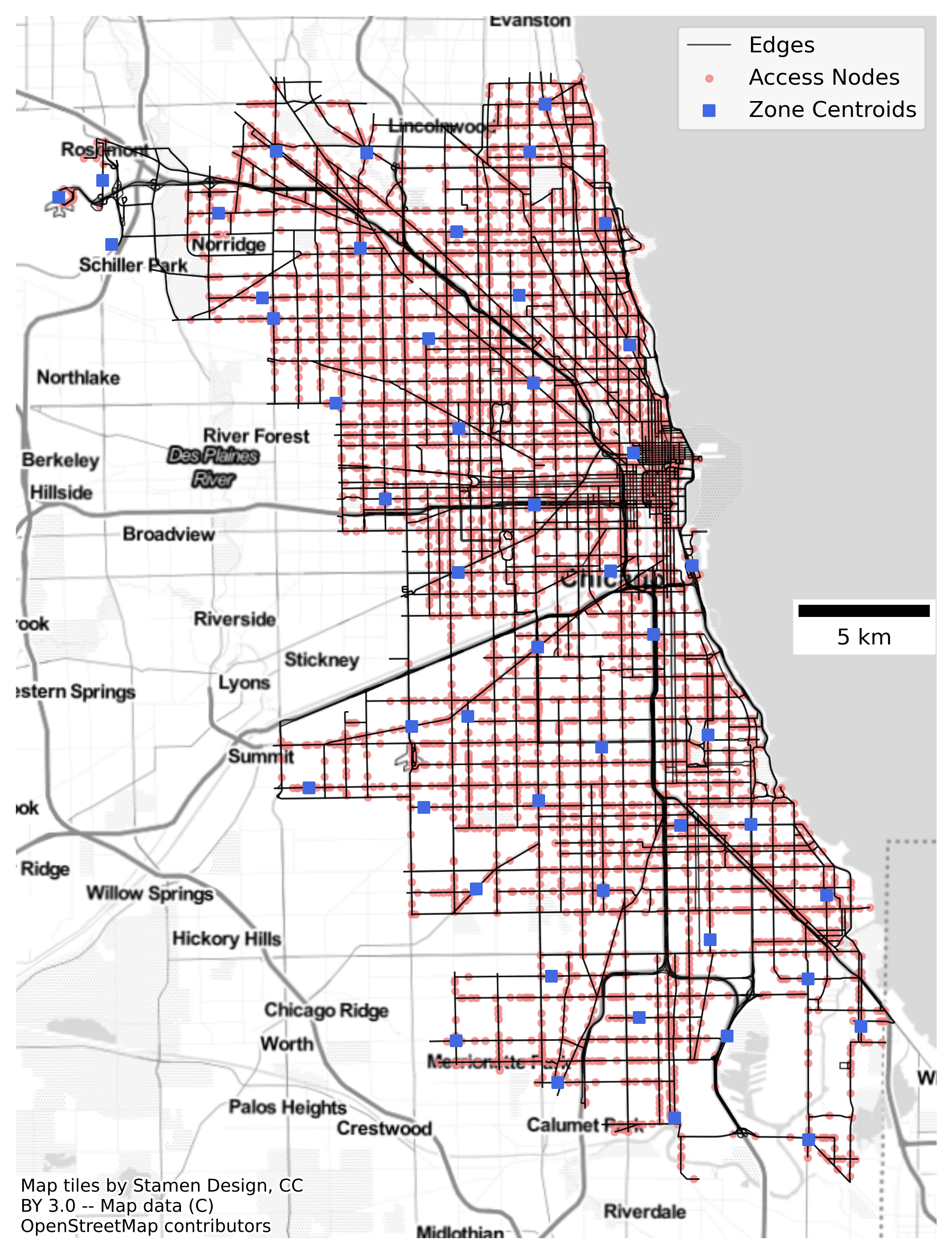}}
  \caption{Chicago network for applied for the case study.}\label{fig:chicago_map}
\end{figure}

The proposed algorithm is evaluated for a case study of Chicago, Illinois. The street network is extracted from OpenStreetMap (OSM) using the python OSMnx package~\citep{Boeing.2017}. To reduce the size of the network, edges labeled as “residential” or “living\_street” are removed from the network, resulting in $12585$ nodes with $27446$ edges. Customers are only allowed to start and end their trip at certain access nodes. Similar to \citep{FlorianDandl.2020} boarding is prohibited on major roads like highways. Therefore, all nodes with adjacent edges not labeled as “primary”, “secondary”, “tertiary” or unlabeled edges are not considered access nodes. The set of access nodes is further reduced by randomly removing access nodes if no other access node can be found within a distance of $300$m. This procedure is repeated until $4000$ access nodes are left, resulting in a small enough number to preprocess travel time tables between those nodes to reduce computational time needed for routing queries. Figure~\ref{fig:chicago_map} shows the resulting network with all access nodes.

Demand for the ODRP service is created using the publicly available TNC data set for Chicago, Illinois~\citep{ChicagoDepartmentofBusinessAffairs&ConsumerProtection.2022}. For this study, TNC trips for 06/07/2023 are used. Trips are removed that start or end outside of the Chicago city boundary. Additionally, presumably faulty data entries and/or round trips are removed, characterized by a trip distance larger than $100$km or lower than $0.1$km, a trip time larger than $5$hours or lower than $60$seconds, and an average speed higher than $130$km/h or lower than $5$km/h. After the filtering process, $127528$ trips remain. Requests are created by choosing a random access node for origin and destination within the reported pick-up and drop-off area. As request time, a random value in second steps is drawn from the reported $15$min start time interval of the trip. To further reduce computational time of the simulations, only a $20$\% subsample of the created requests is used.

To calibrate network travel times, the reported trip duration in the data set is compared to the travel time of the fastest path when considering the maximum allowed speed from the OSM data on each edge. After scaling edge travel times by a factor of $1.62$, the travel times computed by fastest path equal the reported travel times on average.

The operator employs vehicles with capacity $c_v=4$. The maximum waiting time constraint is set to $t_{max}^{wait}=6$min, while a maximum relative detour of $\Delta_{max}^{det}=40\%$ is used. In the base case 300 vehicles are used that serve around 90\% of the demand when repositioning is applied.

Similar to \citep{Wallar.2018}, zones and corresponding centroids are created solving a maximum coverage problem: Let $K_n$ be the set of access nodes reachable from node n within a maximum driving time of $t_{max}^{wait}$. The minimum set of zone centroid nodes that guarantee that each access node is reachable by at least one centroid node within a maximum driving time of $t_{max}^{wait}$ is determined by solving the following ILP:

\begin{align}
    \label{eq:case_1}
    \text{Minimize:} \qquad & \qquad \sum_{n} x_n & \\
    \label{eq:case_2}
    \text{s.t.:} \qquad & \qquad \sum_{\hat{n} \in K_n} x_n \geq 1 & \forall n \\
    \label{eq:case_3}
    & \qquad x_n \in \{0, 1\} & \forall n
\end{align}

$x_n$ are the decision variables that indicate if a node is assigned to be a centroid node. The resulting $48$ zone centroids are shown in Figure~\ref{fig:chicago_map}. Nodes are assigned to the zone of the closest centroid.

Two different methods are tested to forecast future trips:
\begin{enumerate}
    \item Perfect Forecast: From the input request set the number of requests between zone $i$ and $j$ in forecast interval $T$ is used as the Poisson rate $\lambda_{i,j}^T$.
    \item Myopic Forecast: The number of trip requests in the simulation in the past time interval $\{t-\delta_T, t\}$   between zone $i$ and $j$ is used as Poisson rate $\lambda_{i,j}^T$ at time $t$ for each $T$.
\end{enumerate}
It can be assumed that more sophisticated forecast algorithms based on historic trip data should perform at least as good as the myopic forecast, while the perfect forecast acts as an upper bound.   

Repositioning trips are calculated every $\delta_T=900$s. In the base scenario, a forecast horizon of $\mathfrak{H}=2700$s with $N_S=3$ and $\gamma=0.5$ is used. For the comparison algorithms, different parameter variations are tested first and the best performing used in the case study. For the \textit{QT}-Algorithm, $\mu_{QT}=0.7$ and $\mathfrak{H}_{QT}=2700$s is used. $\mu_{Hor}=0.1$ and $\mathfrak{H}_{Hor}=1800$s is used for the \textit{Hor}-Algorithm.

All computations are implemented in Python and conducted on an Intel Xeon Silver processor with $2.10$GHz and $192$GB RAM. Optimization problems are solved with the commercial solver "Gurobi" (\textit{https://www.gurobi.com/}).

\section{Results}

In this section, results of the case study are presented. First, the benefits of rebalancing fleet vehicles are shown. In a second step, the proposed algorithm is compared to other state-of-the-art rebalancing algorithms. Finally, sensitivities to hyperparameters are evaluated.

\begin{figure}
     \centering
     \begin{subfigure}[T]{0.43\textwidth}
         \includegraphics[width=\textwidth]{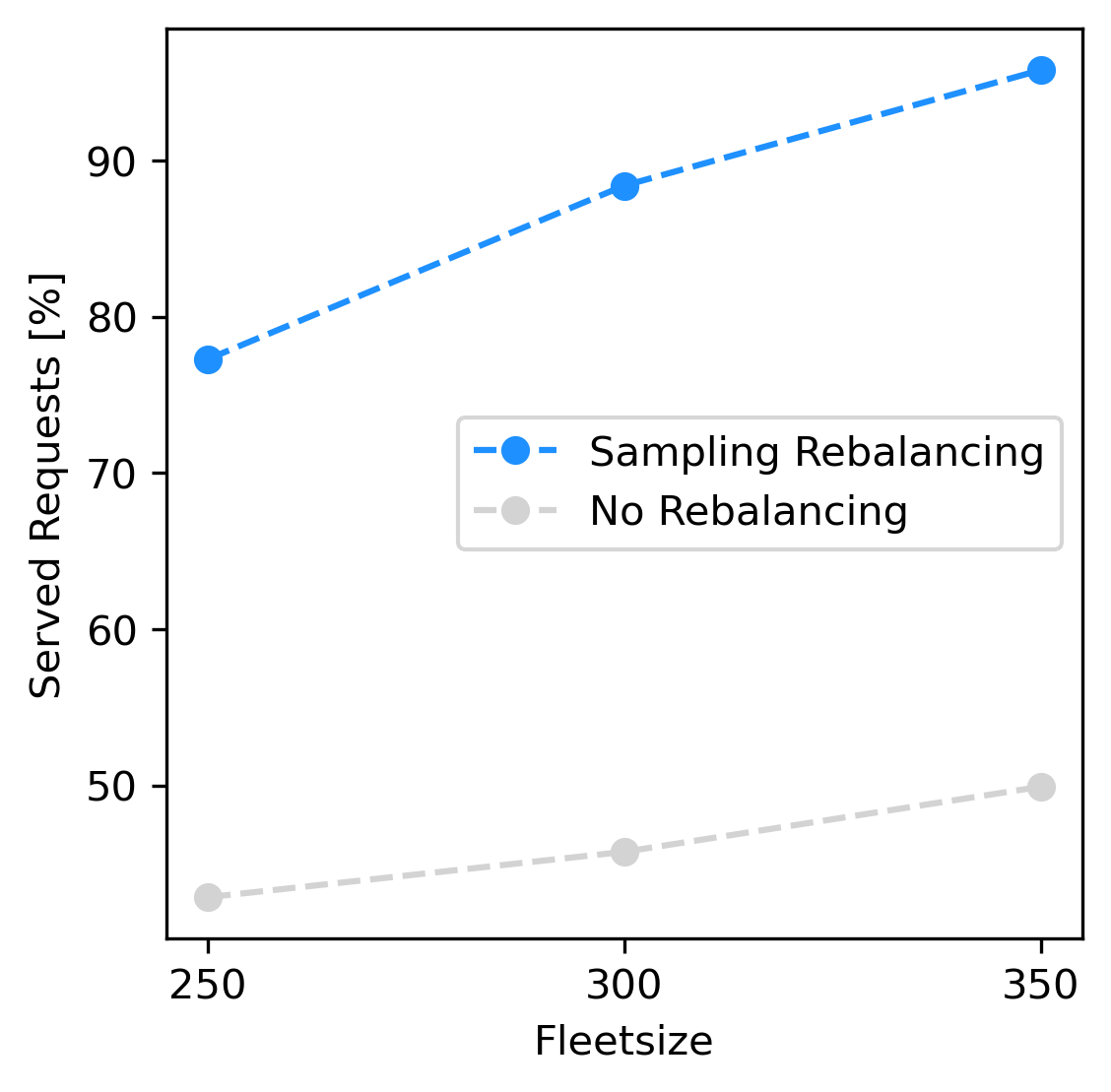}
         \caption{Served Requests for different Fleet Sizes.}
         \label{fig:comp_no_repo_served}
     \end{subfigure}
     \hfill
     \begin{subfigure}[T]{0.43\textwidth}
         \includegraphics[width=\textwidth]{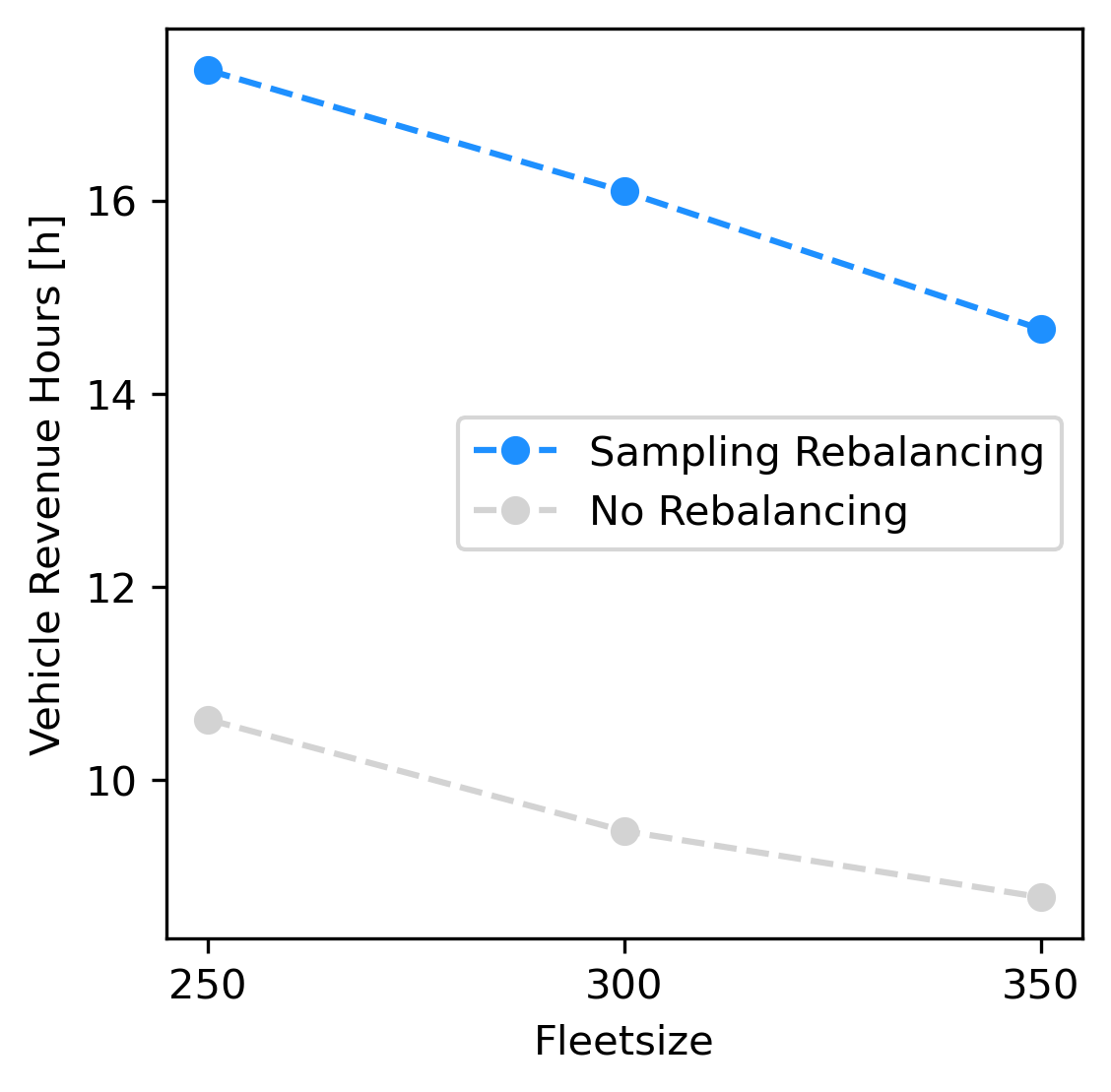}
         \caption{VRH for different Fleet Sizes.}
         \label{fig:comp_no_repo_VRH}
     \end{subfigure}
     \hfill
     \begin{subfigure}[T]{0.43\textwidth}
         \includegraphics[width=\textwidth]{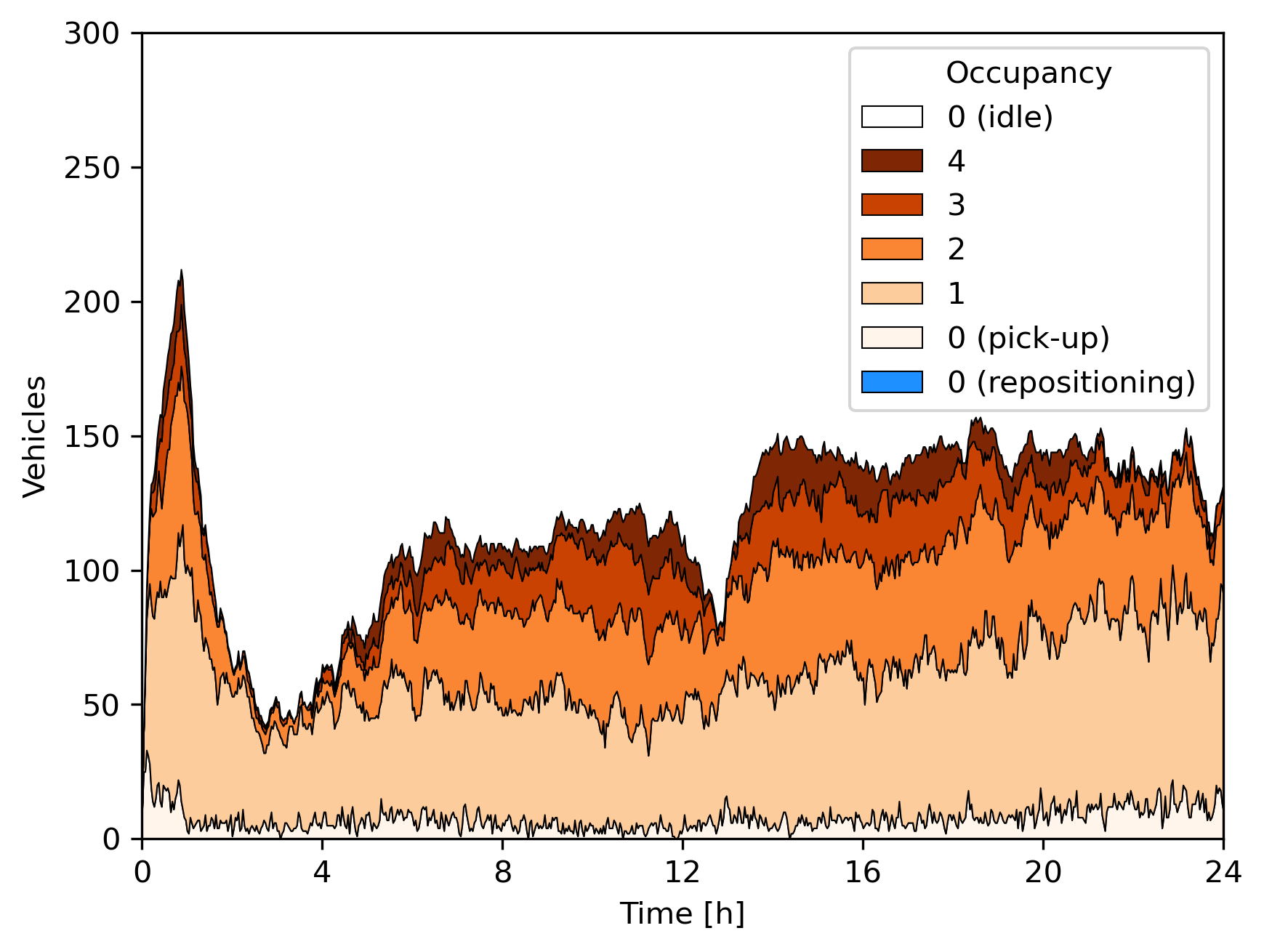}
         \caption{Without rebalancing: Temporal vehicle states of the simulation period.}
         \label{fig:occ_no_rebal}
     \end{subfigure}
     \hfill
     \begin{subfigure}[T]{0.43\textwidth}
         \includegraphics[width=\textwidth]{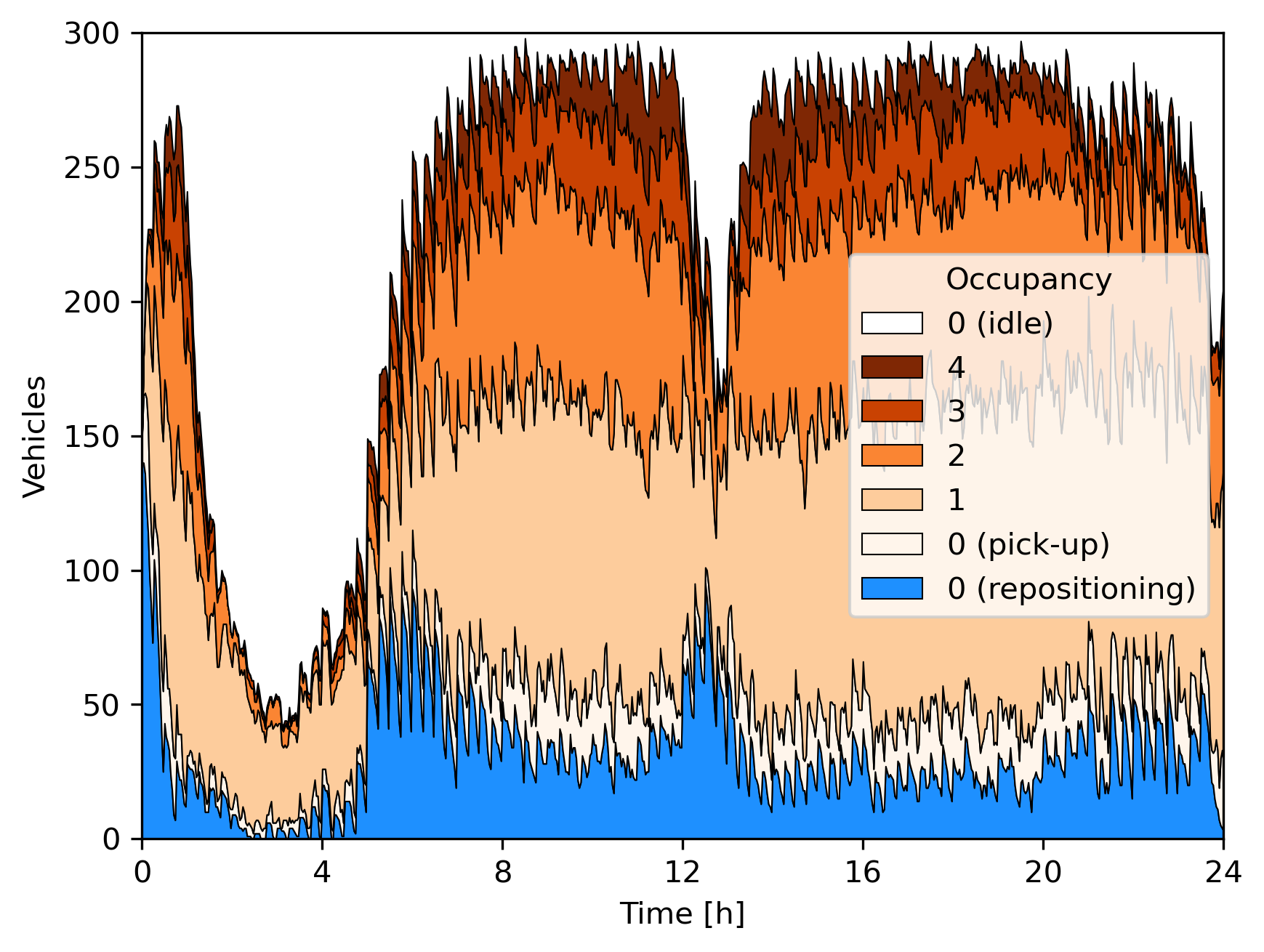}
         \caption{With rebalancing: Temporal vehicle states of the simulation period.}
         \label{fig:occ_with_rebal}
     \end{subfigure}
     \hfill
     \begin{subfigure}[T]{0.43\textwidth}
         \includegraphics[width=\textwidth]{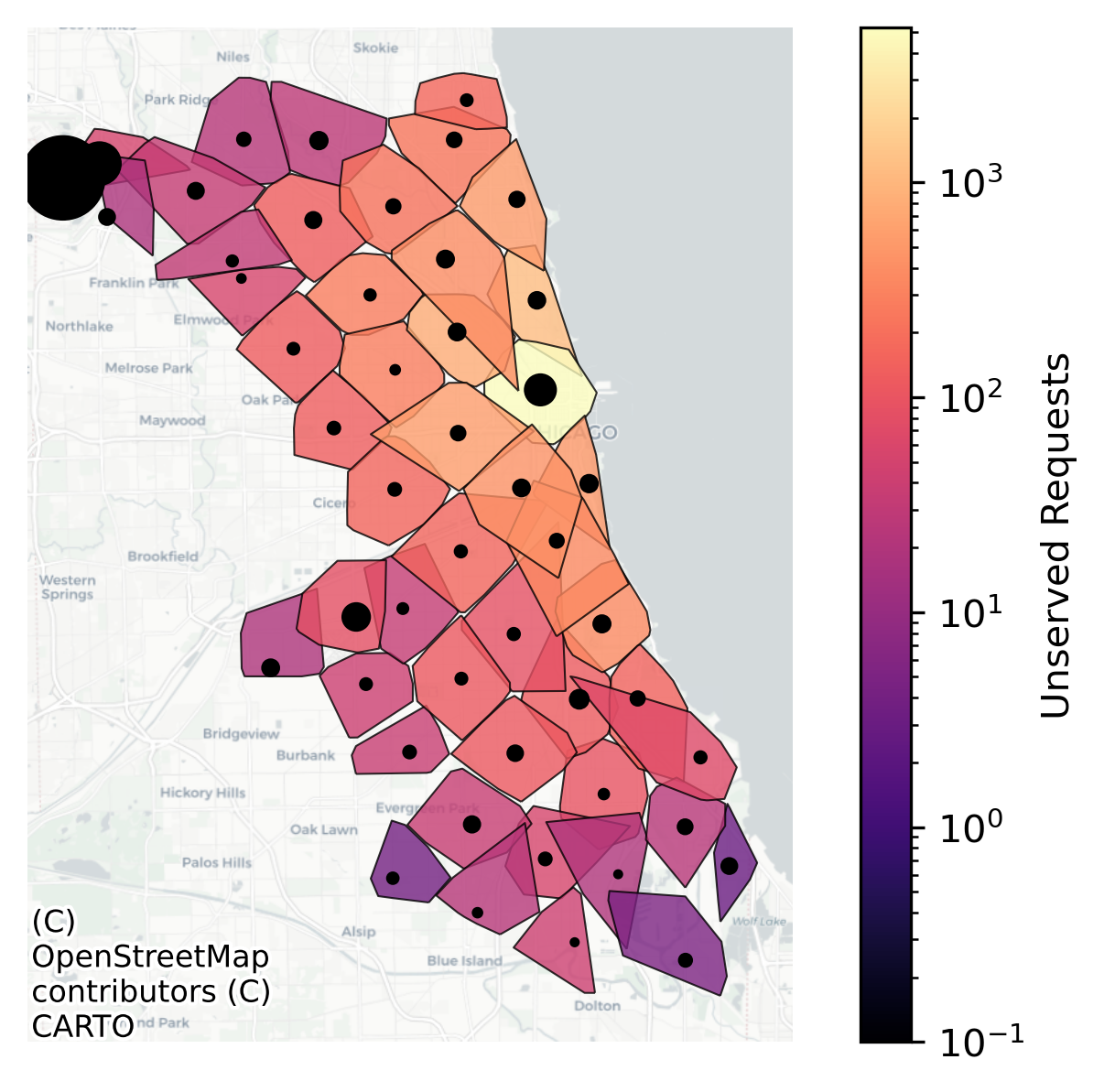}
         \caption{Without rebalancing: Spatial distribution of unserved requests. The size of black circles indicates idle times of fleet vehicles in zones}
         \label{fig:zones_no_rebal}
     \end{subfigure}
     \hfill
     \begin{subfigure}[T]{0.43\textwidth}
         \includegraphics[width=\textwidth]{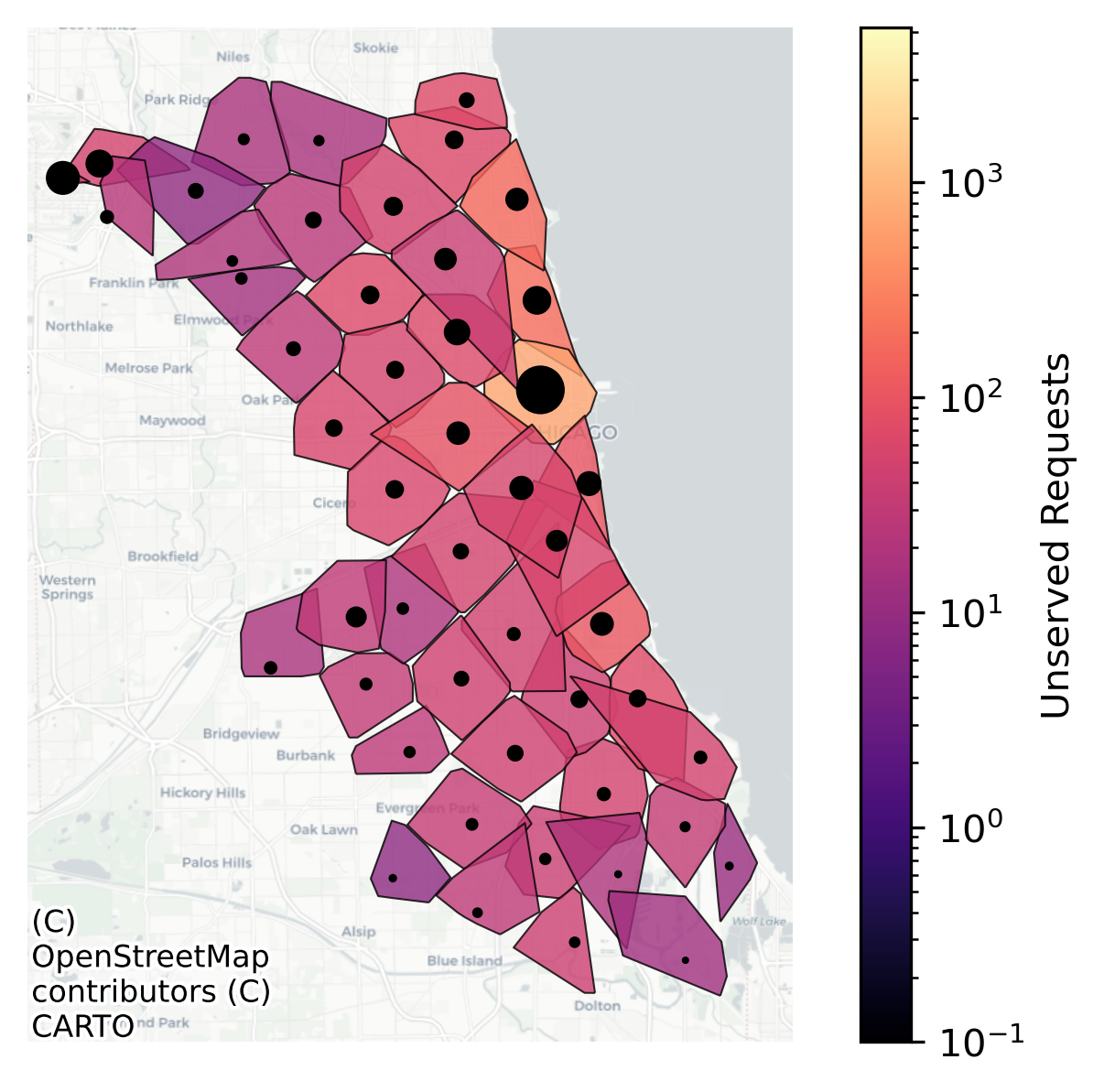}
         \caption{With rebalancing: Spatial distribution of unserved requests. The size of black circles indicates idle times of fleet vehicles in zones.}
         \label{fig:zones_with_rebal}
     \end{subfigure}
        \caption{Comparison of results with and without rebalancing. 300 vehicles are used if not specified.}
        \label{fig:comp_no_repo}
\end{figure}

Figure~\ref{fig:comp_no_repo} shows different evaluations that highlight the comparison between a service that applies the proposed rebalancing algorithm and a service without any rebalancing. Figure~\ref{fig:comp_no_repo_served} shows the number of served requests for three different fleet sizes. When repositioning is applied, 300 vehicles are sufficient to serve 88\% of the daily demand. 93\% of all customers can be served when 350 vehicles are operated. If no rebalancing mechanism is provided, the service rate drops drastically to only around 50\% for 350 vehicles. This drop results from vehicles ending up in network regions with low demand. Vehicles in these regions remain idle until a new customer requests a trip. 

Figure~\ref{fig:comp_no_repo_VRH} shows the average vehicle revenue hours, i.e. the absolute time interval during the day fleet vehicles carry customers and therefore produce revenue for the operator. It can be observed that while vehicles produce revenue for at least 14.5 hours of the day for scenarios with rebalancing, this quantity is reduced to less than 11 hours without rebalancing as a large fraction of vehicles waits for new customers in their current vicinity. 

This effect can also be seen in Figure~\ref{fig:occ_no_rebal} and Figure~\ref{fig:occ_with_rebal} that show the temporal evolution of fleet states during the day for a service with 300 vehicles without and with rebalancing, respectively. By repositioning, the time vehicles spend idle can be reduced significantly, leading to almost full utilization except for times with low demand during night and at noon. Without repositioning, many vehicles stuck in regions with low demand and remain idle, even at times of high demand during the day. Figure~\ref{fig:zones_no_rebal} and Figure~\ref{fig:zones_with_rebal} show the spatial distribution of unserved request on a logarithmic scale. In both scenarios, most requests are rejected in the city center and at the O’hare airport in the northwest corner of the operating area. The absolute number is much lower when rebalancing is applied.  Black circles indicate the time vehicles spend idle in the corresponding zone. Without rebalancing, vehicles especially end up idling at the airport, while with rebalancing idle times are reduced overall and vehicles tend to be located in areas with high demand.

\begin{figure}
     \centering
     \begin{subfigure}[T]{0.43\textwidth}
         \includegraphics[width=\textwidth]{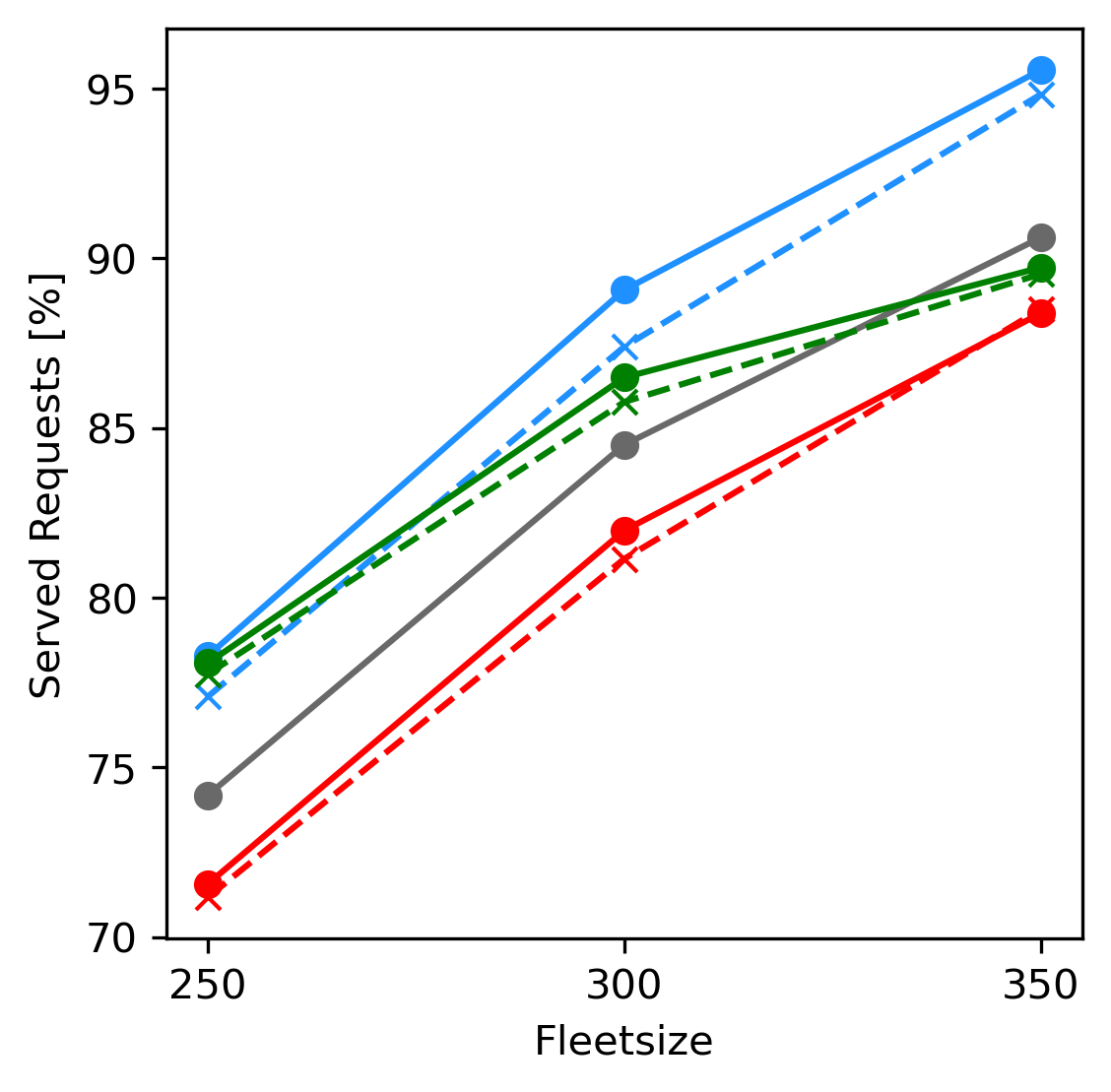}
         \caption{Served Requests.}
         \label{fig:all_served}
     \end{subfigure}
     \hfill
     \begin{subfigure}[T]{0.43\textwidth}
         \includegraphics[width=\textwidth]{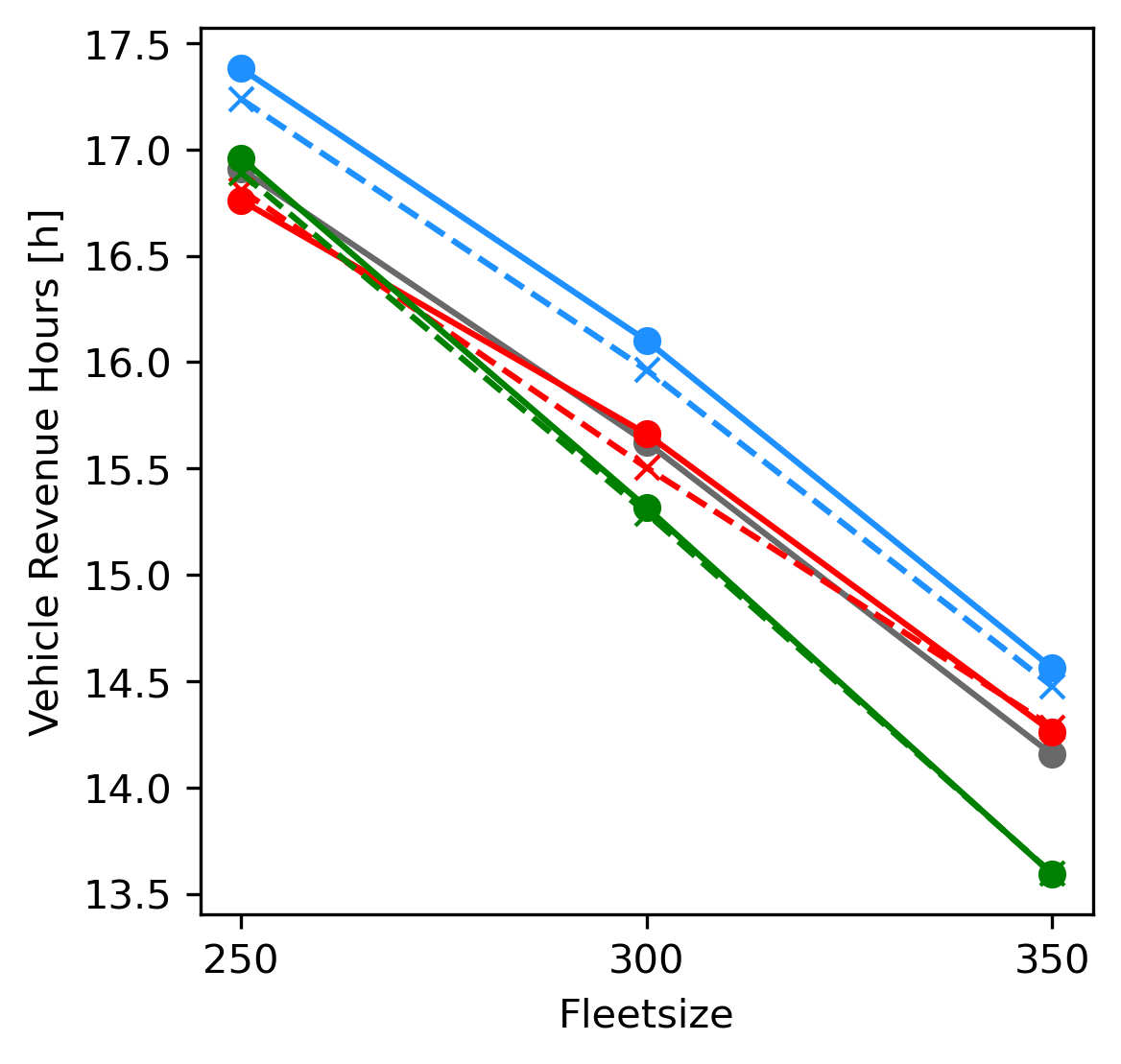}
         \caption{Vehicle Revenue Hours.}
         \label{fig:all_VRH}
     \end{subfigure}
     \hfill
     \begin{subfigure}[T]{0.43\textwidth}
         \includegraphics[width=\textwidth]{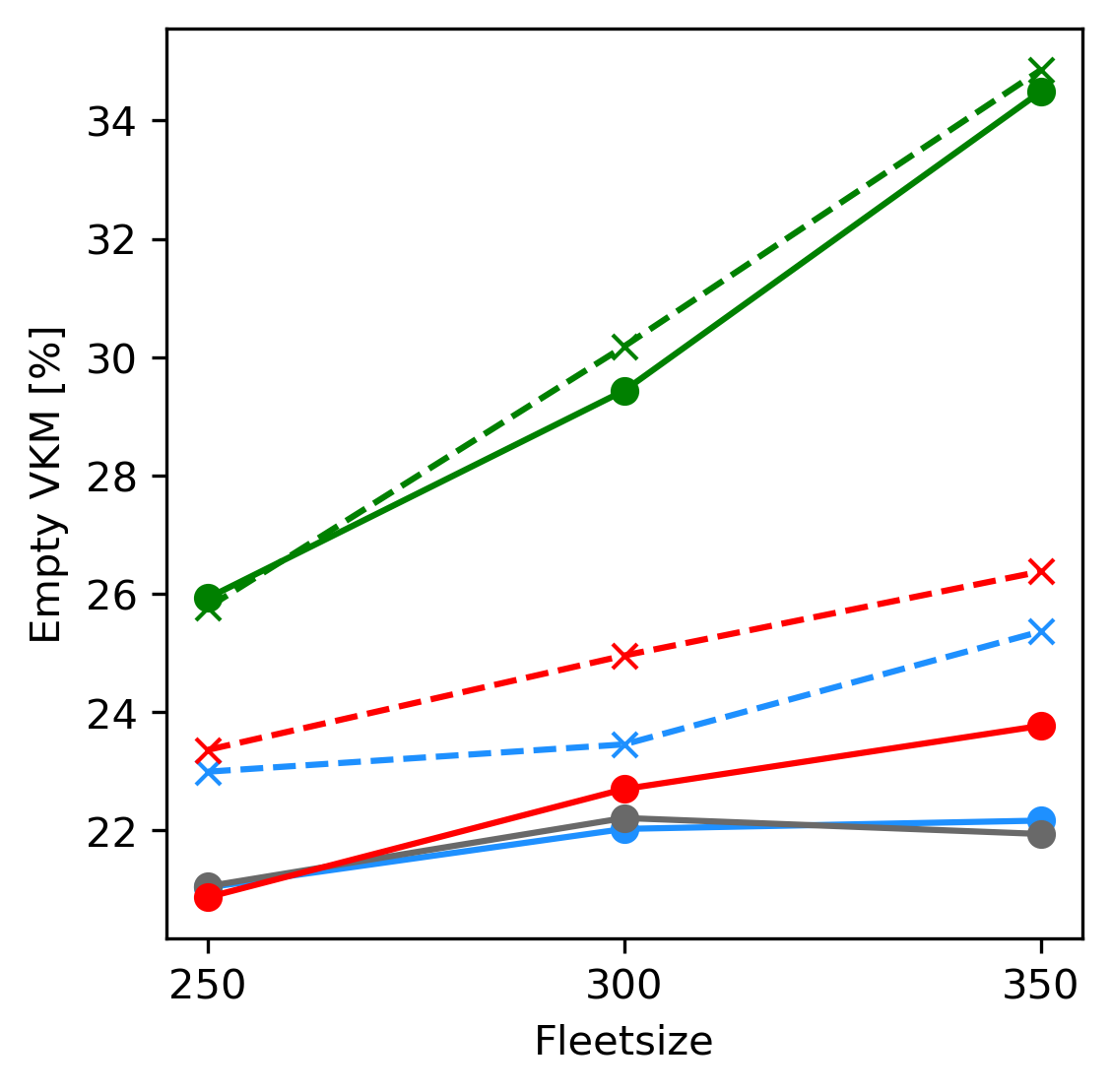}
         \caption{Empty Vehicle Kilometers.}
         \label{fig:all_empty}
     \end{subfigure}
     \hfill
     \begin{subfigure}[T]{0.43\textwidth}
         \includegraphics[width=\textwidth]{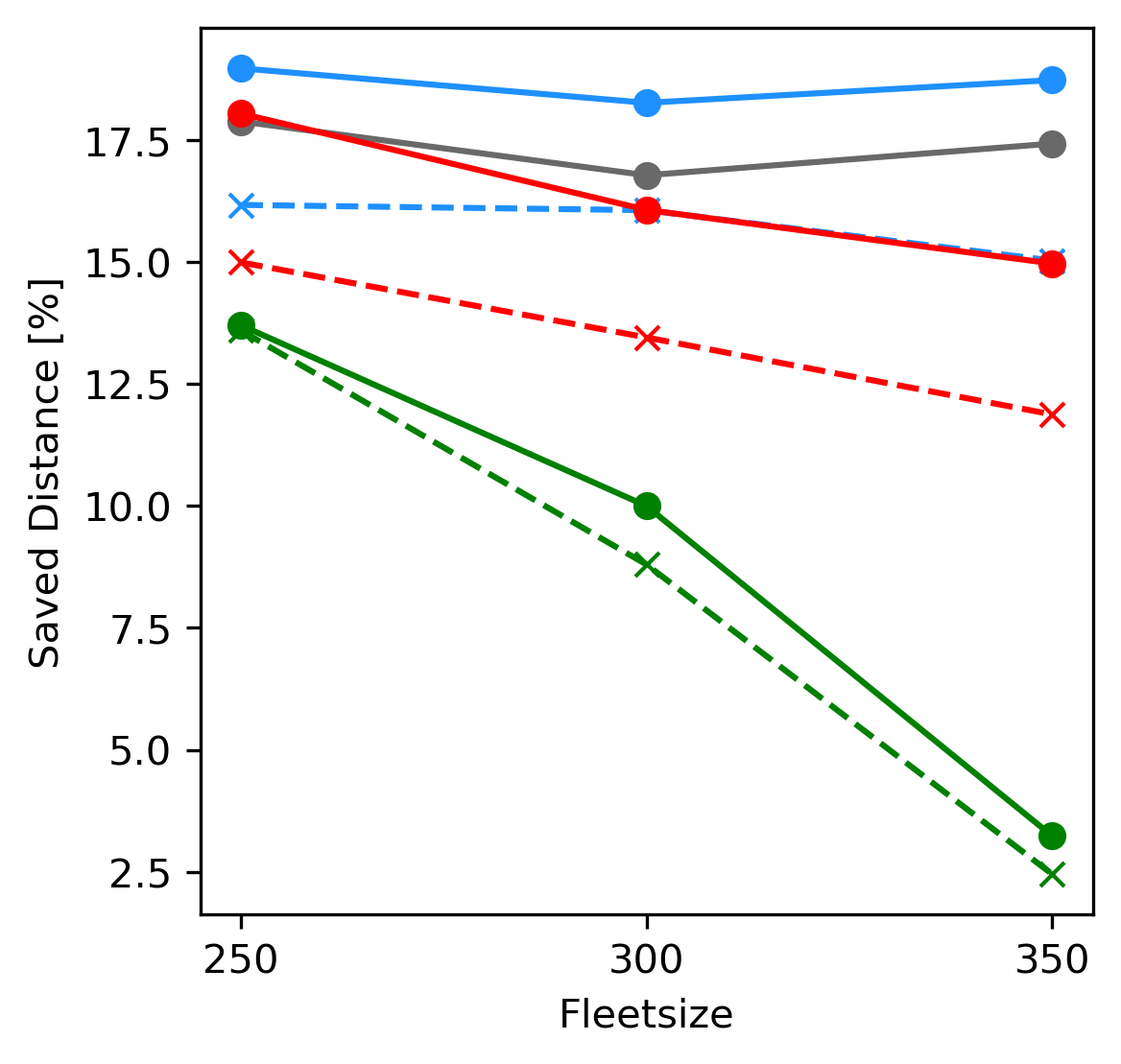}
         \caption{Saved Distance.}
         \label{fig:all_saved}
     \end{subfigure}
     \hfill
     \begin{subfigure}[T]{0.43\textwidth}
         \includegraphics[width=\textwidth]{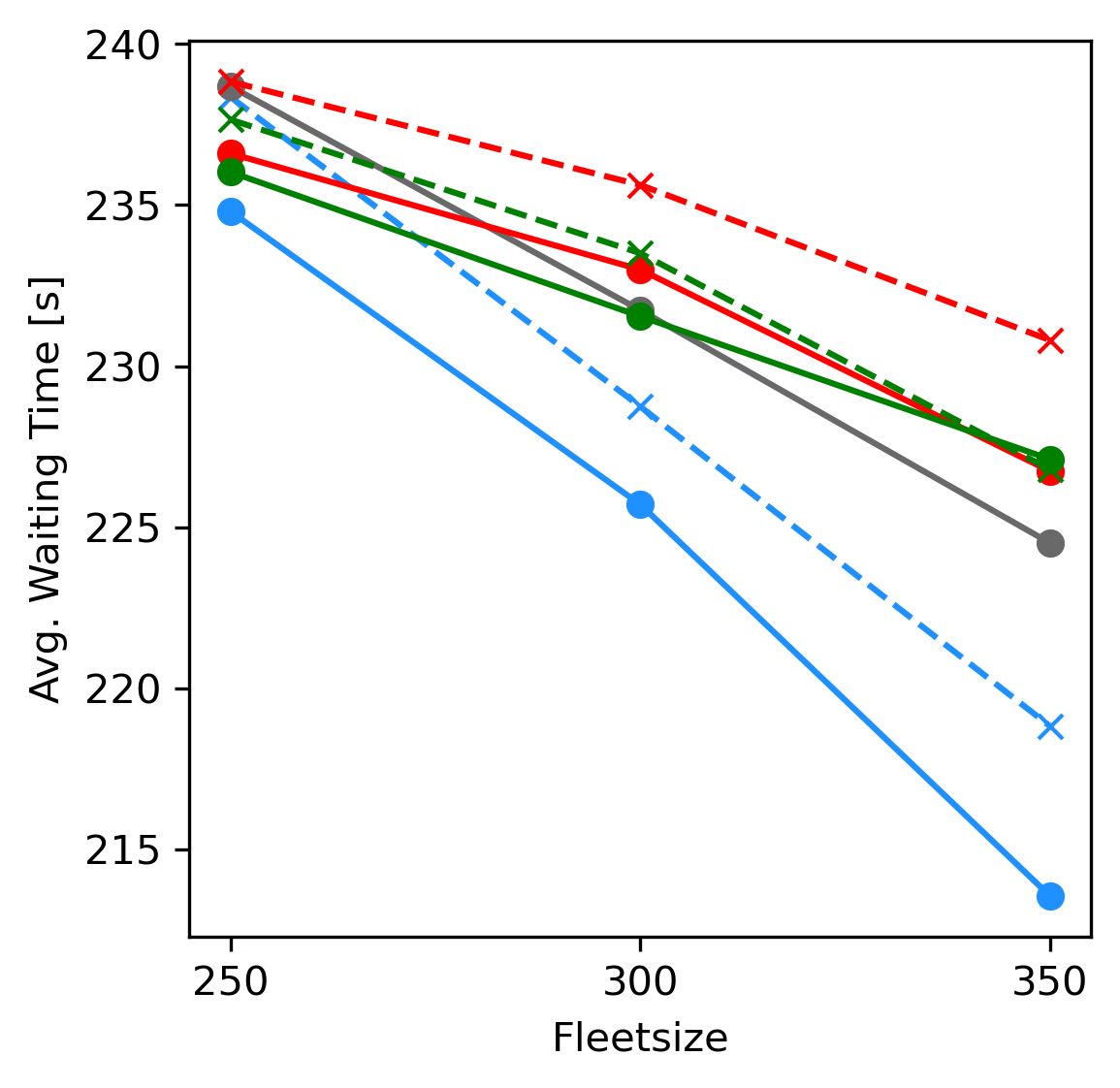}
         \caption{Average Customer Waiting Time.}
         \label{fig:all_wait}
     \end{subfigure}
     \hfill
     \begin{subfigure}[T]{0.35\textwidth}
        \centering
         \includegraphics[width=\textwidth]{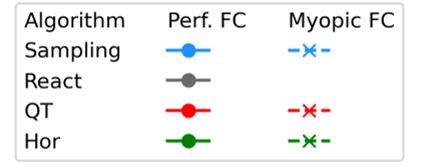}
         %\caption{Legend.}
         \label{fig:all_legend}
     \end{subfigure}
        \caption{Comparison of KPIs with other rebalancing algorithms. As the React-algorithm does not use any forecast, only one graph is shown.}
        \label{fig:comp_all}
\end{figure}

Figure~\ref{fig:comp_all} compares different Key Performance Indicators (KPI) of the ODRP service when different repositioning algorithms and forecast methods are applied. Figure~\ref{fig:all_served} shows the served customers. In comparison with Figure~\ref{fig:comp_no_repo_served} it can be seen, that all approaches outperform a service without rebalancing by far showing the importance of applying rebalancing algorithms for ODRP services. Comparing the different algorithms, the proposed sampling method outperforms all other algorithms except for a fleet size of 250 vehicles and the myopic forecast. In this case, the \textit{Hor} method serves slightly more customers. Nevertheless, the performance of the \textit{Hor}-algorithm degrades for larger fleet size relative to the other algorithms. Even with the myopic forecast, the sampling method produces better results than the React method, which does not apply any forecast at all. This shows that the ODRP service benefits from predicted rebalancing, even for imprecise forecasts. Interestingly, the \textit{QT}-approach performs worse. A reason could be that the introduced scaling parameter $\mu_{QT}$ to scale the demand forecast might not be sufficient to consider pooling in the formulation.

The high service rate directly translates to increased vehicle revenue hours as seen in Figure~\ref{fig:all_VRH}. Vehicles produce revenue for around $30$min longer when rebalanced with the proposed sampling algorithm compared to the other algorithms indicating that vehicles are efficiently repositioned to regions where they are needed. 

Figure~\ref{fig:all_empty} depicts the empty vehicle kilometers of the fleet, which includes rebalancing trips as well as empty pick-up trips. The largest fraction of empty VKM is observed for the \textit{Hor} method indicating an aggressive assignment of rebalancing to idle vehicles, which results in a trade-off compared to the high value of served customers in Figure 4a. The proposed sampling method on the other hand performs well in both KPIs. An imprecise forecast increases empty VKM by around 1-2\% when the myopic forecast is used.

The saved distance in Figure~\ref{fig:all_saved} measures the efficiency of pooling. It is calculated as
\begin{align}
    SD = 1 - \frac{d_{fleet}}{\sum_{i \in R_{served} } d_i } ~,
\end{align}
$d_{fleet}$ is the fleet driven distance, $d_i$ the direct distance of request $i$ travelling from origin to destination and $R_{served}$ the set of served requests. It measures the relative reduction of vehicle kilometers compared to when all customers travel on their own in a private vehicle. Figure~\ref{fig:all_saved} shows that this KPI is positive for all rebalancing methods indicating that pooling overcomes empty vehicle kilometers. Nevertheless, the high empty VKM of \textit{Hor} method leads to the lowest saved distance. Again, if the perfect forecast is applied, the sampling method performs best. With low empty VKM, this method produces the highest pooling efficiency by efficient predictive repositioning. Also saved distance slightly decreases when the myopic forecast is used, but the sampling method still performs well compared to the other algorithms especially considering the high service rate.

Figure~\ref{fig:all_wait} shows average customer waiting times. For all scenarios tested, the sampling method also offers the lowest waiting times to its customers.

\begin{figure}
     \centering
     \begin{subfigure}[T]{0.32\textwidth}
         \includegraphics[width=\textwidth]{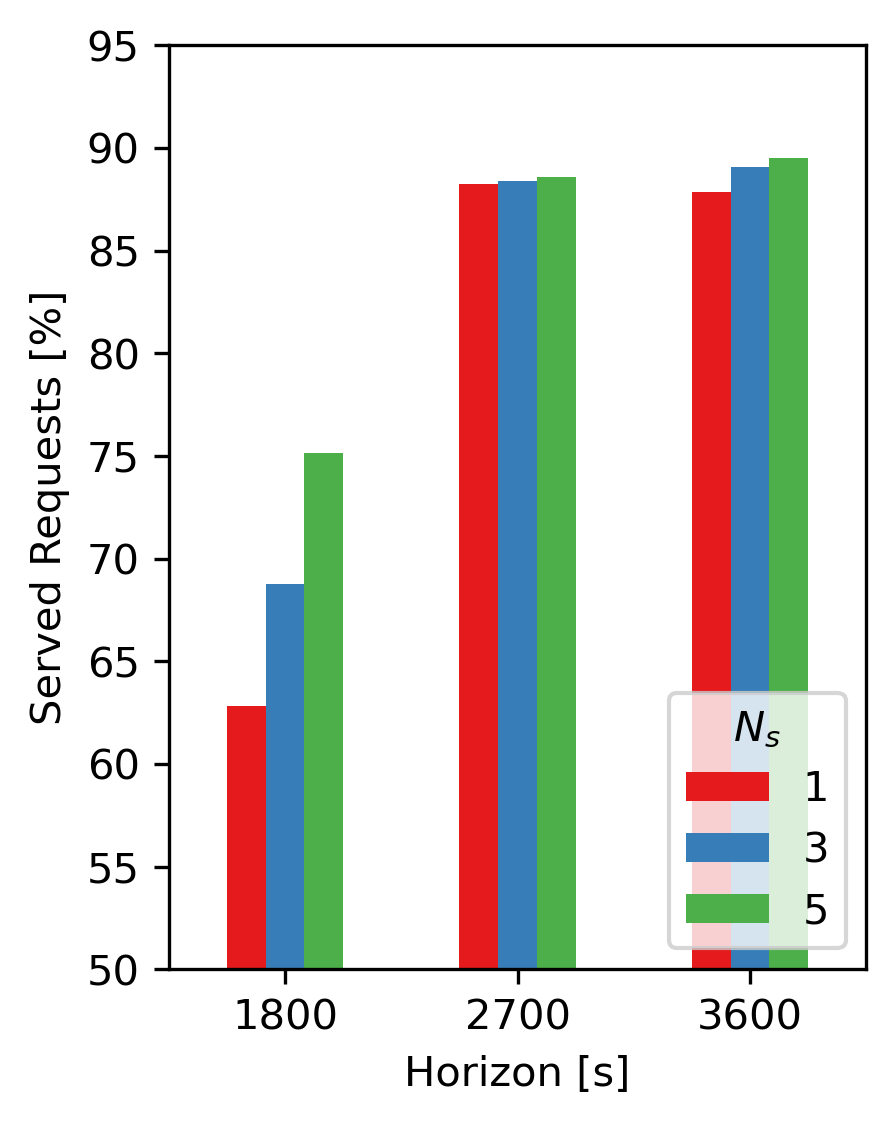}
         \caption{Served Requests.}
         \label{fig:hyper_served}
     \end{subfigure}
     \hfill
     \begin{subfigure}[T]{0.32\textwidth}
         \includegraphics[width=\textwidth]{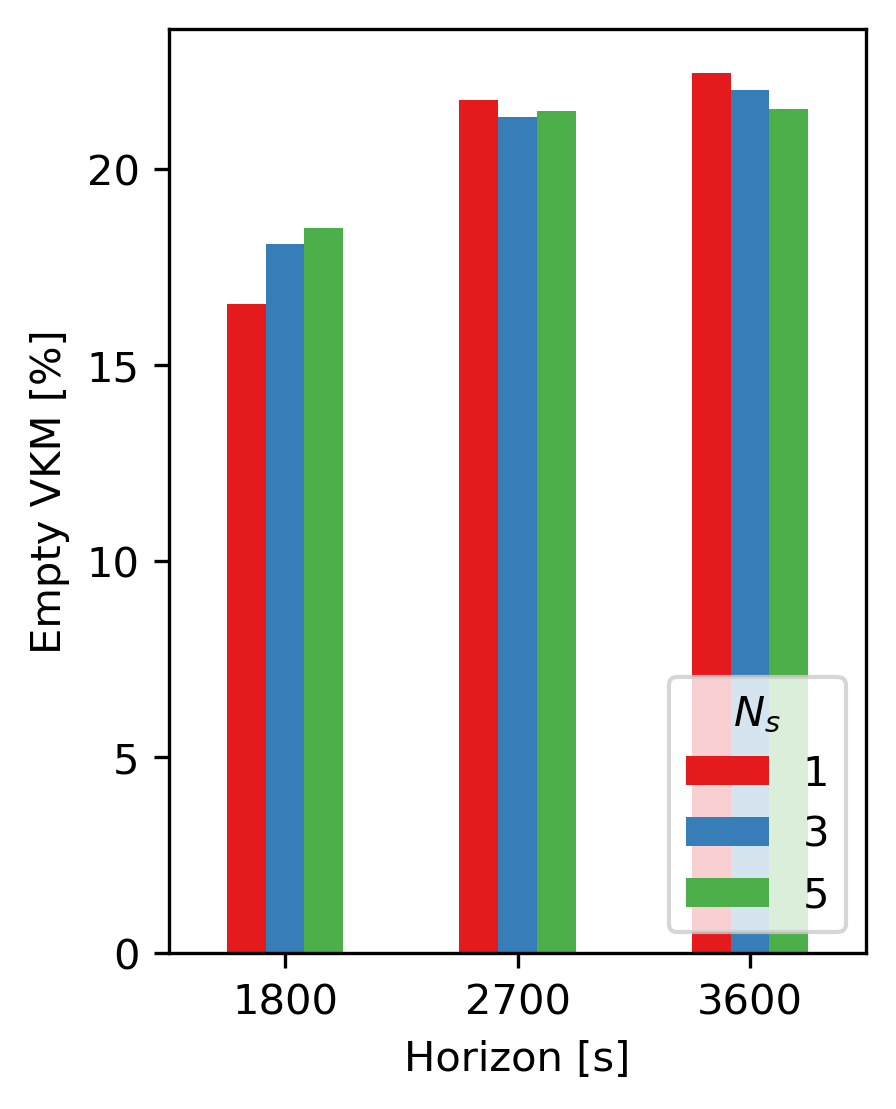}
         \caption{Empty VKM.}
         \label{fig:hyper_empty}
     \end{subfigure}
     \hfill
     \begin{subfigure}[T]{0.32\textwidth}
         \includegraphics[width=\textwidth]{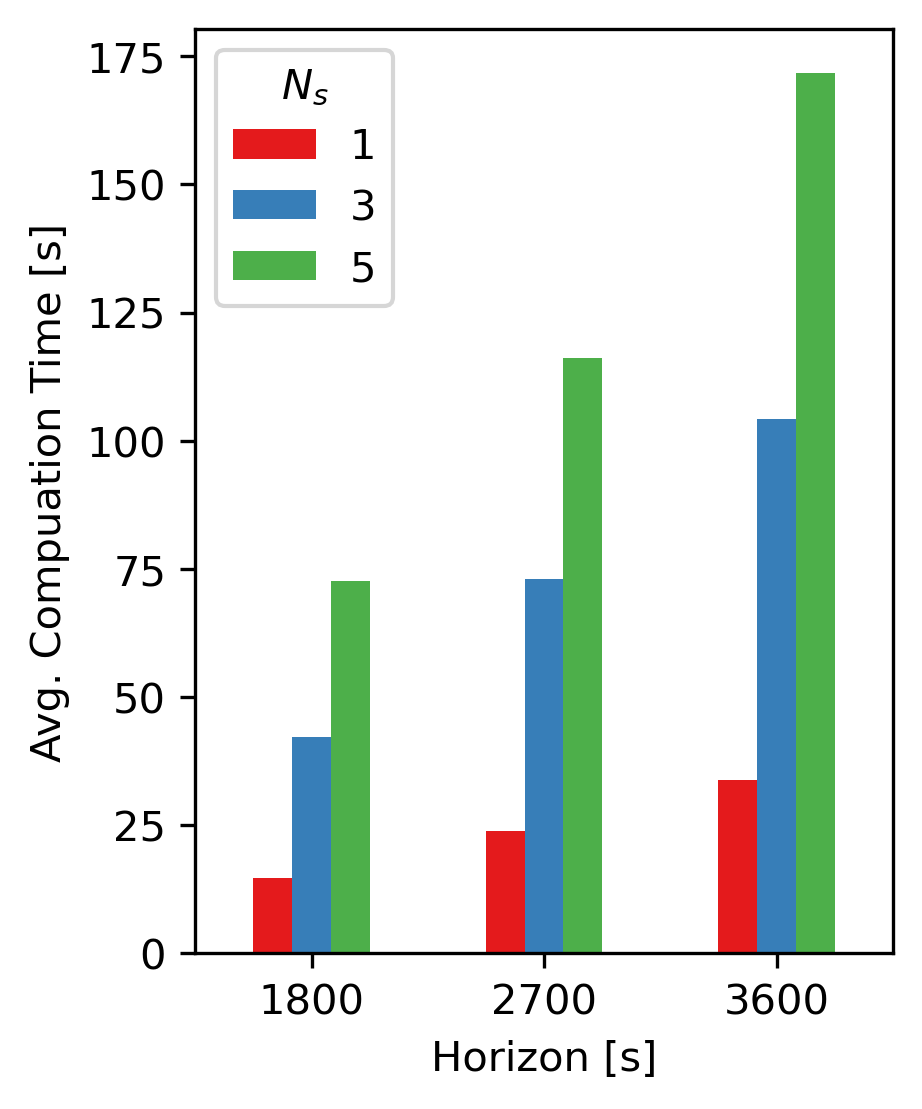}
         \caption{Computational Time.}
         \label{fig:hyper_comp}
     \end{subfigure}
        \caption{Hyperparameter Sensitivity.}
        \label{fig:hyper}
\end{figure}

Finally, Figure~\ref{fig:hyper} shows the sensitivity of hyperparameters $\mathfrak{H}$ (forecast horizon) and $N_S$ (number of samples). It can be observed that $\mathfrak{H}=30$min is not sufficient for the algorithm to achieve its potential. As constraints prohibit vehicles to rebalance over a longer travel time, this horizon likely is not able to cover the whole operating area of Chicago. The effect of the number of samples used in the rebalancing formulation is smaller. The observable trend is that with more samples more requests can be served and less empty VKM is driven, if $\mathfrak{H}$ exceeds $45$min. This can likely be traced back to a better estimation of future supply shortage distributions. Unsurprisingly, the computational time per rebalancing time step increases with $\mathfrak{H}$ and $N_S$. It reaches up to 170s on average for the scenario with $\mathfrak{H}=60$min and $N_S=5$. The computational times of the comparison algorithms are not shown in this figure. As macroscopic formulations are used, these algorithms can be solved within a few seconds. Nevertheless, as the rebalancing algorithm is called every 900s, the sampling algorithm can still be applied in real services. Additionally, all simulations are made in single processing mode. Especially the sampling process can be easily parallelized, which makes up the bulk of the computational time. The ILP to assign tours can be solved within a few seconds.

\section{Summary and Future Work}

This study proposed an algorithm to rebalance idle vehicles to match future demand and supply for an on-demand ride-pooling service.  To estimate future spatio-temporal vehicle supply distributions when trips can be shared, requests are sampled from a demand forecast distribution and vehicle routes are created. An assignment problem is solved to assign vehicle rebalancing trips to maximize expected profit across multiple samples. A case study for Chicago, Illinois showed the huge benefits (e.g. nearly doubling the number of served requests) for the service if a rebalancing algorithm is applied. Also in comparison with other rebalancing algorithms in the literature, the proposed algorithm performs best in increased service rate, pooling efficiency and vehicle revenue hours, and decreased empty vehicle kilometers and customer waiting times. As a tradeoff, the computational time increases but as it is still considerable smaller than the repositioning frequency, real world applications are suitable.

In future work, the rebalancing assignment process will be further refined. For example, the rebalancing targets of vehicles can be set freely instead of aggregated to zone level as information on node level is produced in the sampling process.  Additionally, the sampling process allows incorporating stochasticity and dynamism of network travel times. 

\section{Acknowledgements}

The authors thank the German Federal Ministry of Transport and Digital Infrastructure for providing funding through the project "TEMPUS" with grant number 01MM20008K and fruitful discussion in the project “MINGA” with grant number 45AOV1001K. The authors remain responsible for all findings and opinions presented in the paper. The authors thank Florian Dandl for reviewing the paper and providing valuable comments.

\section{Author Contributions}

The authors confirm contribution to the paper as follows: study conception and design: Roman Engelhardt, Hani S. Mahmassani, Klaus Bogenberger; data collection: Roman Engelhardt; analysis and interpretation of results: Roman Engelhardt, Hani S. Mahmassani, Klaus Bogenberger; draft manuscript preparation: Roman Engelhardt. All authors reviewed the results and approved the final version of the manuscript. 

%Bibliography
\bibliographystyle{plainnat}
\bibliography{bib}

\end{document}